\documentclass[10pt,a4paper,oneside]{article}

\usepackage[english]{babel}
\usepackage{lmodern}
\usepackage{amsmath} 
\usepackage{amssymb} 
\usepackage{graphicx}
\usepackage{gensymb}
\usepackage{longtable}
\usepackage[text={13cm,20cm},centering]{geometry}
\usepackage{icomma} 

\newcommand{\myfootnote}[1]{%
\renewcommand{\thefootnote}{}%
\footnotetext{#1}%
\renewcommand{\thefootnote}{\arabic{footnote}}%
}

\begin{document}

\noindent
{\Large{\textbf{\center Analysis of the latitudinal data of Eratosthenes and Hipparchus}}}\myfootnote{}
\\\\\\
Christian Marx$^*$\myfootnote{$^*$Christian Marx, Gropiusstra{\ss}e 6, 13357 Berlin, Germany; e-mail: ch.marx@gmx.net.}\myfootnote{\;\;Revision submitted to MEMOCS on 11 August 2014.}
\\\\\\
\noindent
\textit{Abstract:}
The handed down latitudinal data ascribed to Eratosthenes and Hipparchus are composed and each tested for consistency by means of adjustment theory.
For detected inconsistencies new explanations are given concerning the origin of the data.
Several inconsistent data can be ascribed to Strabo.
Differences in Hipparchus' data can often be explained by the different types and precision of the data.
Gross errors in Eratosthenes' data are explained by their origination from the lengths of sea routes.
From Eratosthenes' data concerning Thule a numerical value for Eratosthenes' obliquity of the ecliptic is deduced.

%========================================================================
\section{Introduction}
%========================================================================

A precise specification of positions on the earth surface became possible in ancient geography by the introduction of reference systems and physical quantities for the description of positions.
Eratosthenes (ca.\ 276--194 BC), founder of mathematical geography, introduced a grid of non-equidistant parallels and meridians for the positions of selected places.
In his ``Geography'' he described the inhabited world by means of distance data and expressed his latitudinal data probably by means of meridian arc lengths.
The astronomer and mathematician Hipparchus (ca.\ 190--120 BC) probably introduced the division of the full circle into 360\degree\ into Greek astronomy and geography (e.g., \cite[p.\ 149]{dic60}).
He transferred the concept of ecliptical longitude and latitude for the specification of star positions to the terrestrial sphere.
Hipparchus' essential geographical work is his treatise ``Against the `Geography' of Eratosthenes'', wherein he discussed the works of Eratosthenes and gave a compilation of latitudes and equivalent astronomical quantities for several locations.
Later Ptolemy (ca.\ 100--170 AD) used Hipparchus' concept and introduced a geographical coordinate system for his position data in his ``Geography'' (\textit{Geographike Hyphegesis}, GH), which differs from today's system only by its zero meridian.

The mentioned works of Eratosthenes and Hipparchus are handed down only in fragments, mainly by Strabo's (ca.\ 63 BC -- 23 AD) ``Geography'' (G; see \cite{jon17}, \cite{rad11}).
The geographical fragments of Eratosthenes and Hipparchus were compiled and commented on by \cite{ber80} and \cite{rol10} as well as by \linebreak
\cite{ber69} and \cite{dic60}, respectively.
In particular the latitudinal data of the fragments are given partly redundantly and with differing numerical values.
It is uncertain, whether all the data originate from Eratosthenes or Hipparchus, respectively (see also \cite[p.\ 36]{rol10}).
Therefore, an investigation of their consistency is indicated.
The aim of this contribution is to carry out such an investigation conjointly for all considered data so that it is ensured and shown that all relations between the data are integrated.
For this purpose, the data are composed in systems of equations; when solving these systems appropriately, the inconsistencies of the data become evident (Sections \ref{sec:erat}, \ref{sec:hipp}).
For inconsistencies new explanations are given.
The actual accuracy of the investigated ancient data is not subject of this contribution.
Some grossly erroneous data of Eratosthenes, however, are explained by their origination from the lengths of sea routes (Section \ref{sec:searoutes}).

Among Eratosthenes' latitudinal data there is a distance concerning Thule, the place visited by the geographer and astronomer Pytheas during his expedition to Great Britain in about 330 BC.
From Eratosthenes' and Pytheas' information concerning Thule a numerical value for Eratosthenes' obliquity of the ecliptic is deduced (Section \ref{sec:obliquity}).
A localisation of Thule based on new considerations of Pytheas' sea route and of the lengths of the nights in Thule is to be found in the appendix.

%========================================================================
\section{Eratosthenes' latitudinal data} \label{sec:erat}
%========================================================================

The latitudinal data of Eratosthenes considered in the following are based on the fragments given by \cite{rol10} (the following translations are taken from it).
The investigations are limited to the locations in the western \textit{Oikoumene} (the inhabited world known to the Greeks and Romans), in particular those in connection with Eratosthenes' prime meridian through Rhodes, because only these data are partly redundant.
The considered data originate from Strabo's ``Geography'', Pliny's ``Natural History'' (NH; see \cite{bos55}) and Cleomedes' ``Caelestia'' (C).
Figure \ref{fig:meridian} shows some of the locations.

Strabo and possibly Eratosthenes expressed latitudes and latitudinal differences by means of meridian arc lengths ($b$ hereinafter and $b_0$ if with respect to the equator) in the measurement unit stadium (st).
Eratosthenes introduced the value
\begin{equation} \label{eqn:C}
C=252,000\,\text{st}
\end{equation}
for the circumference of the earth (e.g., G~II.5.7) so that for a meridian arc holds true
\begin{equation} \label{eqn:degree_erat}
1\degree \mathrel{\widehat{=}} C / 360 = 700\,\text{st} \; .
\end{equation}

The considered information presumably originating from Eratosthenes and the corresponding fragments (F) and sources are given in Table \ref{tab:eratdata}.\footnote{The symbols $\lesssim$ and $\gtrsim$ stand for ``somewhat smaller'' and ``somewhat larger''.}
The F-numbers correspond to \cite{rol10}.
The data occurring repeatedly within one fragment are listed and used once only.
In addition, consecutive numbers were introduced, which are separated from the F-number by a dot.
If it follows from the textual source that two locations have the same latitude, $b$ is set to 0.
Figure \ref{fig:erat} visualises the data in a graph.
An edge is set between two locations if at least one $b$ exists for them.

%------------------------------------------------------------------------
\subsection{Notes on the data} \label{sec:erat_notes}
%------------------------------------------------------------------------

The southern limit of the inhabited world is constituted by the Cinnamon country (F34/G~II.5.7).
For the northern limit, Eratosthenes gives two locations of differing latitudes: the ``northern regions'' (F34/G~II.5.9) and Thule (F35/G~I.4.2).

The Cinnamon country corresponds to the stretch of coast between Cape Guardafui and the straits of Bab-el-Mandeb (cf.\ \cite[p.\ 171]{dic60}).
The information of F30.1, F30.2 (G~II.5.6) and F34.11 (G~II.5.7) does not refer directly to the Cinnamon country but to the southern limit of the inhabited world.
Their identicalness results from G~II.5.7.

The data on the Borysthenes, the Dnieper River\footnote{In antiquity, there was disagreement on the location of the Borysthenes, cf.\ Pliny, NH~IV.83. For instance,  Ptolemy (GH~III.5) probably confuses the Borysthenes with the Hypanis and locates the Borysthenes at the Southern Bug (cf.\ \cite[pp.\ 50, 53]{markle12}).}, refer to its mouth into the Black Sea.

The latitude of the Hellespont (Dardanelles) presumably corresponds to that of Eratosthenes' parallel through Lysimachia (near Bolay{\i}r, Gallipoli peninsula), which is mentioned in G~II.5.40 (likewise \cite[p.\ 155]{ber80}).

F34 (G~II.5.7):
Strabo says that $b_0$ of the tropic of Cancer corresponds to $\frac{4}{60}$ of $C$ (i.e.\ $C/15=16,800$\,st), that the tropic goes through Syene and that $b_0$ of Syene is 16,800\,st.
The latter is applied here (F34.5).
Eratosthenes uses $\frac{1}{15}$ of the full circle for the obliquity of the ecliptic $\varepsilon$ in this case, i.e.\
\begin{equation} \label{eqn:epsr}
\varepsilon_\text{r} = 24\degree \; ,
\end{equation}
which was a common value in early Greek geography (cf.\ \cite[pp.\ 733--734]{neu75}).

F34.13 (G~II.5.8):
This fragment is introduced in addition to \cite{rol10} (see Section \ref{sec:erat_results}) but not used in the following test of consistency.

F36.1, F36.2 (G~II.5.42), F47.1 (G~II.1.3):
In G~II.5.42 Strabo deals with Hipparchus' data on the regions in the neighbourhood of the Borysthenes and the southern parts of Lake Maeotis (Sea of Azov) and he states: ``Eratosthenes says that these regions are a little more than 23,000 stadia [F36.1] from Mero\"{e}, since it is 18,000 stadia [F36.2] to the Hellespont and then 5,000 [F36.3] to Borysthenes.''
\cite[p.\ 155]{rol10} derives that the ``\ldots mouth of the Borysthenes is somewhat over 23,000 stadia from Mero\"{e}.''
From the text, however, follows that $b$ of Mero\"{e} -- Borysthenes is $(18,000+5,000)\,\text{st}=23,000\,\text{st}$.
Thus, the text suggests that Eratosthenes differentiated between the latitude of the mouth of the Borysthenes and the latitude of the mentioned regions, which are situated further north than the mouth.\footnote{In fact, however, the southern parts of Lake Maeotis are further south than the Borysthenes.}
Another interpretation results from F34.1/35.1 (Mero\"{e} -- Alexandria) and F35.2 (Alexandria -- Hellespont), which yield $b=18,100$\,st for Mero\"{e} -- Hellespont so that $b$ of Mero\"{e} -- Borysthenes is 23,100\,st.
Possibly, Eratosthenes specified $b=23,100$\,st for the mouth of the Borysthenes and its neighbouring regions and Strabo described this value by ``a little more than 23,000\,st'' and erroneously used 18,000\,st instead of 18,100\,st in his statement.
Moreover, in G~II.1.3 Strabo states: ``From Mero\"{e} to the Hellespont is no more than 18,000 stadia [F47.1] \ldots''.\footnote{I thank one of the referees for his reference to this text passage.}
Since, however, he compares this value with the distance from India to the Bactrians,
it is certainly only a rough value.
Owing to the differences in the information, the values $23,000$\,st and $18,000$\,st are not used in the following calculational test of consistency.

F40 (G~II.1.20), F41 (NH~II.185):
According to F40, Eratosthenes agrees closely with Philo (Ptolemaic officer, see \cite[p.\ 157]{rol10}) that in Mero\"{e} the sun is at the zenith 45 days before the summer solstice.
In order to test this, the dates of the year were determined for 350--250 BC when the sun altitude $a$ was maximal in Mero\"{e} ($\phi=16\degree56'$).\footnote{\label{foo:altitude}The time of the summer solstice can be determined according to \cite[pp.\ 165--167]{mee91}.
For a location of longitude $\lambda$ and latitude $\phi$ and for a given time $t$, the sun altitude $a$ can be determined by the following calculation steps (formulas see \cite[pp.\ 84, 88--89, 135, 151--3]{mee91}):
obliquity of the ecliptic: $\varepsilon(t)$;
mean anomaly of the sun: $M(t)$;
mean ecliptic longitude of the sun: $L_0(t)$;
equation of center: $C(t,M)$;
true ecliptic longitude: $\Theta(L_0, C)$;
right ascension of the sun: $\alpha(\Theta, \varepsilon)$;
declination of the sun: $\delta(\Theta, \varepsilon)$;
GMST: $\theta_0(t)$;
hour angle: $H(\lambda, \alpha, \theta_0)$;
$a(\phi, \delta, H)$.
(The software-implemented calculation was tested by means of a comparison with the results of the online calculator by \cite{cor13}.)}
As a result, the sun reached its maximal $a\approx90\degree$ 45 or 46 days before as well as 46 or 45 days after the day of the summer solstice.
Hence, the information of F40 is probably based on an accurate observation.
Eratosthenes' $b$-data yield $b_0=11,800$\,st for Mero\"{e} (see Section \ref{sec:erat_results}), which corresponds to $\phi=16\degree51'$ in good agreement with the actual $\phi$.
According to F41, the shadows fall to the south 45 days before and after the summer solstice in the country of the Trogodytes.
That corresponds to the information of F40 on Mero\"{e}.
It is not known whether Eratosthenes derived latitudes from the information of F40 and F41.
At least, however, it can be assumed that Eratosthenes believed the Trogodytes and Mero\"{e} to be located at the same latitude.
Thus, $b=0$ is introduced here for Meroe -- Trogodytes only.

F128.1 (G~II.5.24):
The distance Alexandria -- Rhodes is not explicitly indicated as a latitudinal difference, but it was found by ``\ldots\ using the shadow of a gnomon \ldots'' so that it is considered to be a $b$-value.

FM6 (C~I.7):
This fragment originates from Eratosthenes' work ``On the Measurement of the Earth'', see \cite[pp.\ 263--267]{rol10}.
According to this, Syene is located at the tropic of Cancer; thus, following F34, $b_0=16800$\,st is applied here.

%--------------------------------------------------------------------------
\subsection{Test of consistency} \label{sec:erat_results}
%--------------------------------------------------------------------------

\begin{figure}[p]
\centering
\includegraphics[width=8cm]{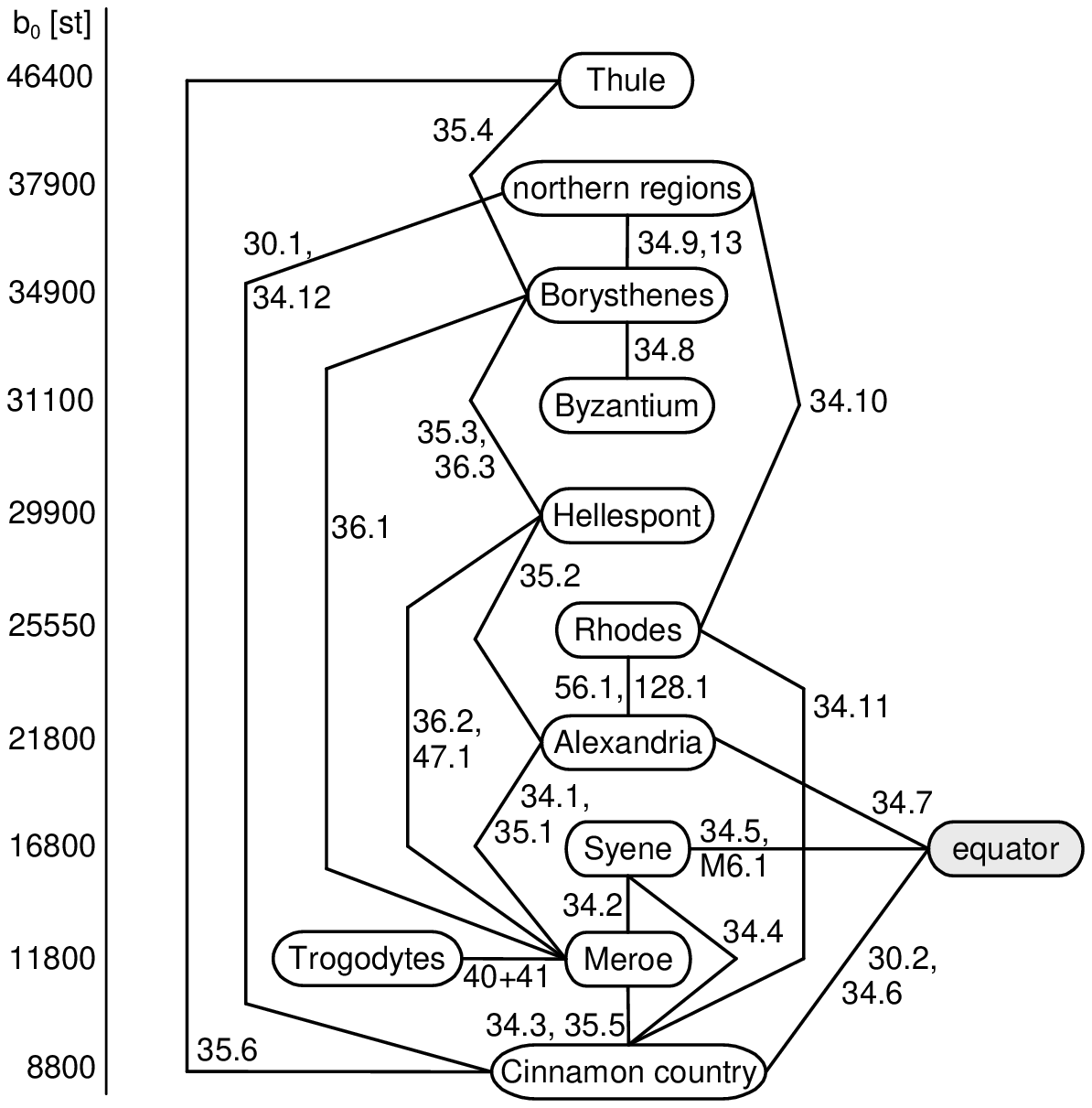}
\caption{Graph of the latitudinal data of the western \textit{Oikoumene} ascribed to Eratosthenes; the vertical order of the locations gives the meridian arc length $b_0$ from the equator (not drawn to scale)}
\label{fig:erat}
\end{figure}

A test of the consistency of the data attributed to Eratosthenes is carried out simultaneously for all data by means of a formation of an equation system for the data.
For each given latitudinal data $b_i^{j,k}$, the equation $b_i^{j,k} = b_{0}^{k} - b_{0}^{j}$ is introduced, where $b_i^{j,k}$ is the $i$-th given meridian arc length between the parallels of the $j$-th and $k$-th location and $b_{0}^{j}$ and $b_{0}^{k}$ are the meridian arc lengths between these locations and the equator.
The latter are the unknown quantities of the equation system, they refer to the 11 locations in Figure \ref{fig:erat}.
If a $j$ refers to the equator, $b_{0}^{j}$ is not an unknown but has the value 0.
If a $b_i^{j,k}$ is specified by an inequality $b>x$ or $b<x$, it is replaced by $b=x$ for the computation.

Since there are redundant data which are inconsistent among each other, the numerical solution of the equation system is not directly possible.
In order to achieve a consistent equation system, an unknown correction $v_i$ (residual) is introduced for each data $b_i^{j,k}$:
\begin{equation} \label{eqn:obs}
b_i^{j,k} + v_i = b_{0}^{k} - b_{0}^{j} \; , \quad  i = 1 \ldots n
\end{equation}
($n$ is the number of data), as it is usual in adjustment theory.
System (\ref{eqn:obs}) can be solved by means of the minimisation of an object function $S$ of the corrections $v_i$.
In the present case the object function
\begin{equation} \label{eqn:min}
S = \sum_{i=1}^{n}{|v_i|} \rightarrow \min
\end{equation}
of the $L_1$-norm adjustment is appropriate because it is a resistant estimation method and therefore able to reveal inconsistencies of the data (see, e.g., \cite{mar13a}; data errors are manifested in large values of the $v_i$).
For the $b$-values which occur multiply in Table \ref{tab:eratdata}, multiple equations are introduced so that their influence is increased in the adjustment.
The $L_1$-norm adjustment is numerically solved here by the simplex algorithm by \cite{bar74} (BR-algorithm).

The imprecise data of F30.1, F34.12 and F56.1 expressed by inequalities are not included in the adjustment but also obtain corrections $v_i$ based on the determined unknowns.
There remain $n=22$ data for the adjustment.
The solution of the BR-algorithm yields 18 $v_i$ being 0; the belonging $b$-data are therefore consistent among each other.
The other $v_i\neq0$ are considered in the following.
It shows that the solution for the $b_0$ can be regarded as to be in accord with Eratosthenes' original data, with the exception of the ``northern regions'' (see below).
The $b_0$ are given in Figure \ref{fig:erat}.

F30.1, F34.12; Cinnamon country -- ``northern regions''; $b<30,000$\,st; $v=100$\,st:
$b<30,000$\,st is contradictory to $v>0$.
If this is Eratosthenes' information, probably not all $b$-data of F34.3, F34.1/35.1, F35.2, F35.3/36.3 and F34.9 (Cinnamon country -- Mero\"{e} -- Alexandria -- Hellespont -- Borysthenes -- ``northern regions'') originate from Eratosthenes because they yield 30,100\,st.
One explanation is that $b=4000$\,st for Borysthenes -- ``northern regions'' (F34.9/G~II.5.9) is not from Eratosthenes but from Strabo.
This $b$-value is already given in the preceding section G~II.5.8: ``For, so far as science is concerned, it is sufficient to assume that, just as it was appropriate in the case of the southern regions to fix a limit of the habitable world by proceeding three thousand stadia south of Mero\"{e} [\ldots], so in this case too we must reckon not more than three thousand stadia [F34.13] north of Britain [i.e.\ Borysthenes], or only a little more, say, four thousand stadia'' (\cite[p.\ 445]{jon17}).
(The latitude of Britain equals that of the Borysthenes according to G~II.5.8.)
The text suggests that the mentioned value of 3,000\,st (F34.13) may originate from Eratosthenes and the value of 4,000\,st may be an alteration by Strabo.
By means of 3,000\,st, the considered extent of the inhabited world is 29,000\,st, which fulfills the condition $b<30,000$\,st of F30.1 and F34.12, and $b_0$ of the ``northern regions'' is 37,900\,st.

F34.10; Rhodes -- ``northern regions''; $b=12,700$\,st; $v=650$\,st:
$b$ is contradictory to the value 13,350\,st which follows from F128.1 (Alexandria -- Rhodes) and F35.2, F35.3/36.3, F34.9 (Alexandria -- Hellespont -- Borysthenes -- ``northern regions'').
Possibly, $b=12,700$\,st originates from Strabo.
12,700\,st less 4,000\,st for the part Borysthenes -- ``northern regions'' (F34.9) yields 8700\,st for the part Rhodes -- Borysthenes, but F128.1, F35.2 and F35.3/36.3 yield 9,350\,st.
The value 8,700\,st, however, nearly corresponds to Hipparchus' value 8,600\,st (cf.\ G~II.1.12, II.5.41) so that Strabo possibly chose it following Hipparchus and used it for $b$ of F34.10.

F34.11; Cinnamon country -- Rhodes; $b=16,600$\,st; $v=150$\,st:
$b$ is contradictory to the sum 16,750\,st of F34.3, F34.1/35.1, F128.1 (Cinnamon country -- Mero\"{e} -- Alexandria -- Rhodes).
Possibly, $b=16,600$\,st is Strabo's sum, which is not based on 3,750\,st for the part Alexandria -- Rhodes (F128.1) but on 3,600\,st which is given by Hipparchus (cf.\ G~II.5.39 and Section \ref{sec:hipp_results}).

F35.5; Cinnamon country -- Mero\"{e}; $b=3,400$\,st; $v=-400$\,st:
$b$ is contradictory to $b=3,000$\,st of F34.3, which equals Hipparchus' $b$ in G~II.5.35.
Strabo says in F35: ``\ldots\ if we add 3,400 [F35.5] more beyond Mero\"{e}, so that we include the Egyptian island, the Kinnamomophoroi [Cinnamon country], and Taprobane, we have 38,000 stadia [F35.6].''
Hence, the reason for an alteration of $b$ by Strabo could be the extent of the Egyptian island and/or of Taprobane.
In F53 (G~II.5.14) Strabo states that the Cinnamon country, Taprobane and the Island of the Egyptians are situated on the same parallel, but in view of the spatial extent of these three locations this can only be an approximate piece of information (e.g., Eratosthenes' estimate of the latitudinal extent of Taprobane is 7,000\,st according to F76/NH~VI.81).
Strabo possibly introduced 3,400\,st in order to obtain the round value of 38,000\,st (F35.6) for the latitudinal extent of the inhabited world from its southern limit to Thule.

F35.6; Cinnamon country -- Thule; $b=38,000$\,st; $v=-400$\,st:
$b$ equals the sum of the other $b$-data of F35 but it is 400\,st too large with regard to the sum based on Eratosthenes' presumable $b=3,000$\,st (F34.3) for Cinnamon country -- Mero\"{e}, see F35.5.
Thus, $b=38,000$\,st is probably only a round sum for the extent of the inhabited world given by Strabo.

F36.1; Mero\"{e} -- Borysthenes; $b\gtrsim 23,000$\,st; $v=100$\,st:
Since the information on $b$ is consistent with $v>0$ and the small $v$, F36.1 can be regarded as consistent.

F36.2, F47.1; Mero\"{e} -- Hellespont; $b=18,000$\,st; $v=100$\,st:
$b$ is contradictory to the sum $b=18,100$\,st of F34.1/35.1 and F35.2 (Mero\"{e} -- Alexandria -- Hellespont).
Strabo ascribes F35.1 and F35.2 explicitly to Eratosthenes.
The value 18,000\,st possibly originates from Strabo, see Section \ref{sec:erat_notes}.

F56.1; Alexandria -- Rhodes; $b\lesssim 4,000$\,st; $v=-250$\,st:
The information on $b$ is in accord with $v<0$ and $v$ is acceptable because $4,000$\,st is probably a roughly rounded value; hence, F56.1 can be considered to be consistent.

%--------------------------------------------------------------------------
\subsection{Sea routes in Eratosthenes' latitudinal data} \label{sec:searoutes}
%--------------------------------------------------------------------------

The two southernmost $b$ of Cinnamon country -- Mero\"{e} (F34.3/35.5) and of Mero\"{e} -- Alexandria (F35.1) are (almost) correct.
$b=3,000$\,st of F34.3 and $b=3,400$\,st of F35.5 are converted by means of (\ref{eqn:degree_erat}) 4\degree17$'$ and 4\degree51$'$, respectively, and actually 4\degree43$'$ (based on the central latitude 12\degree13$'$ for the Cinnamon country).
$b$ of F35.1 is converted as well as actually 14\degree17$'$.

In contrast, the subsequent latitudinal differences Alexandria -- Hellespont of F35.2 and Hellespont -- Borysthenes of F35.3/36.3 show large errors.
$b=8,100$\,st of F35.2 is $11\degree34'$ and actually 9\degree23$'$; the error is $2\degree11'\mathrel{\widehat{=}}1,528$\,st.
$b=5,000$\,st of F35.3/36.3 is $7\degree09'$ and actually 6\degree01$'$; the error is $1\degree08'\mathrel{\widehat{=}}793$\,st.

\begin{figure}[p]
\centering
\includegraphics[width=8cm]{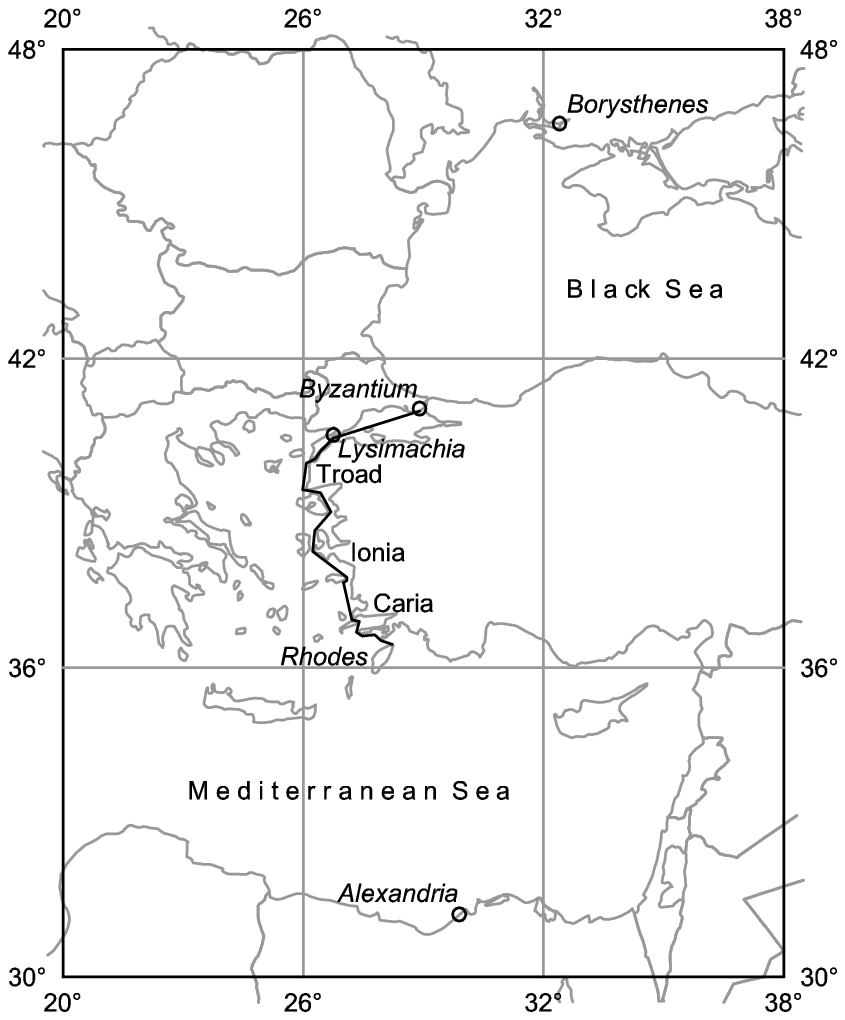}
\caption{Places located on Eratosthenes' prime meridian (\textit{italic}) and supposable sea routes underlying Eratosthenes' data (\textit{lines})}
\label{fig:meridian}
\end{figure}

Both erroneous latitudinal differences are explicable by Eratosthenes' conception that the prime meridian through Rhodes also runs through Mero\"{e}, Alexandria, Caria, Byzantium and (the mouth of) the Borysthenes (cf.\ G~I.4.1, II.1.12, II.1.40).
Figure \ref{fig:meridian} shows the position of the concerning locations.
Moreover, Strabo says that it is generally agreed that the sea route Alexandria -- Borysthenes is a straight line (G~II.5.7).
Consequently, it is likely that the latitudinal differences Alexandria -- Hellespont and Hellespont -- Borysthenes are based on the lengths of the sea routes, which were supposed to take course along the meridian (also assumed by \cite[p.\ 640]{bun79}, \cite[p.\ 152]{rol10}).
This is considered more in detail in the following.
The lengths of the sea routes could have been derived from journey times and estimates of the speed, which was a usual procedure according to GH~I.9.4, I.17.6.

F35.2, Alexandria -- Hellespont:
At least from its part Alexandria -- Rhodes it is known that Eratosthenes had information on the length of the sea route from navigators.
For this part Eratosthenes gives his own $b=3,750$\,st (F128.1) and additionally the two lengths 4,000\,st and 5,000\,st for the corresponding sea route based on the assumptions of navigators (F128/G~II.5.24).
The large difference between both lengths shows the large uncertainty of such information.
If for the latitudinal difference Alexandria -- Rhodes Eratosthenes' 3,750\,st are used, then $(8,100-3,750)\,\text{st}=4,350\,\text{st}$ remain for the rest of distance F35.2, i.e.\ for Rhodes -- Hellespont.
Figure \ref{fig:meridian} shows a possible sea route from Rhodes to Hellespont (Lysimachia).
It has a length of ca.\ 650\,km.
In order to convert this into stadia, not (\ref{eqn:degree_erat}) is used (which only applies to $b$) but a conventional stadium length.
The Egyptian stadium of 157.5\,m (cf.\ \cite[p.\ 33]{dil85}) is chosen\footnote{\label{foo:rasat}For the Egyptian stadium there is some evidence.
Strabo says (G~X.4.5) that the sea voyage from Cyrenaea to Criumetopon/Cape Krio (cf.\ \cite[IV.20, fn.\ 4]{bos55}) in Crete takes two days and nights and that Eratosthenes gave 2,000\,st for this distance.
Pliny also treats that distance (NH~IV.20) and gives more precisely Phycus/Cape Rasat (cf.\ \cite[IV.20, fn.\ 20]{bos55}) instead of Cyrenaea.
The distance Cape Rasat -- Cape Krio is 313\,km, which yields 156.5\,km per day.
Eratosthenes' distance is probably based on the definition: $1\,\text{day-and-night-seafaring} \mathrel{\widehat{=}} 1,000\,\text{st}$, which was also assessed by Theophilus and Marinus (GH~I.9.4).
Hence, journey times were converted into distances by means of this definition; from it and the distance Cape Rasat -- Cape Krio follows $1\,\text{st} \approx 156.5\,\text{m}$.
This is in accordance with the Egyptian stadium of 157.5\,m so that this stadium possibly underlies the aforesaid definition.}, which yields $650\,\text{km} = 4,127\,\text{st}$ in good agreement with the ancient value of 4,350\,st.

F35.3/36.3, Hellespont -- Borysthenes:
$b$ of the part Byzantium -- Borysthenes is 3,800\,st according to F34.8 (G~II.5.8; this value is ca.\ 100\,st less than the actual value).
For the rest of distance F35.3, i.e.\ for Hellespont -- Byzantium,  $(5,000-3,800)\,\text{st}=1,200\,\text{st}$ remain.
The assumed sea route Hellespont -- Byzantium shown in Figure \ref{fig:meridian} has a length of $195\,\text{km} = 1,238\,\text{st}$ in good agreement with the ancient value of 1,200\,st.

%--------------------------------------------------------------------------
\subsection{Eratosthenes' obliquity of the ecliptic} \label{sec:obliquity}
%--------------------------------------------------------------------------

Strabo states (F34/G~II.5.8) that according to Pytheas in Thule the arctic circle coincidences with the tropic of Cancer.
The arctic circle delimits the region of the circumpolar stars in the sky, which do not set (see, e.g., \cite[p.\ 165]{dic60}).
Thus, its declination is
\begin{equation} \label{eqn:arctic}
\delta_\text{a} = 90\degree - \phi \; ,
\end{equation}
where $\phi$ is the latitude of the observer.
Strabo's information means that $\delta_\text{a}$ equals the obliquity of the ecliptic $\varepsilon$.
Hence, for $\phi$ of Thule $\phi=90\degree-\varepsilon$ holds true so that Thule is situated at the northern polar circle.
There the sun does not set at the summer solstice, which corresponds to Pliny's information (NH~IV.30) that at the summer solstice there are no nights in Thule.
$\varepsilon$ was 23\degree44$'$ at the time of Pytheas' voyage\footnote{\label{foo:eps}Computed according to \cite[p.\ 135]{mee91}.} so that the northern polar circle was at $\phi=66\degree16'$.
At Eratosthenes' time $\varepsilon$ was 23\degree43$'$ so that $\phi=66\degree17'$.

The computation of Section \ref{sec:erat_results} yields $b_0=46,400$\,st for Thule.
This result is composed of the following $b$-data:\\
1. equator -- Cinnamon Country: $8,800$\,st (F30.2, F34.6);\\
2. Cinnamon Country -- Mero\"{e}: $3,000$\,st (F34.3, the value occurs twice in F34);\\
3. Mero\"{e} -- Alexandria: $10,000$\,st (F34.1, F35.1);\\
4. Alexandria -- Hellespont: $8,100$\,st (F35.2);\\
5. Hellespont -- Borysthenes: $5,000$\,st (F35.3, F36.3);\\
6. Borysthenes -- Thule: $11,500$\,st (F35.4).\\
The 2nd $b$-value equals the difference between F34.4 (Cinnamon country -- Syene) and F34.2 (Mero\"{e} -- Syene).
The sum of 13,000\,st of the 2nd and 3rd $b$-value (i.e. Cinnamon country -- Alexandria) equals the difference between F34.7 (Alexandria) and the 1st $b$-value.
The sum of the 3rd, 4th and 5th $b$-value (i.e. Mero\"{e} -- Borysthenes) amounts to 23,100\,st and is confirmed by F36.1, where $\gtrsim$ 23,000\,st is given.
Hence, considering the 3rd and 5th $b$-value to be correct, also the 4th $b$-value is confirmed.
The 6th $b$-value was probably calculated from the $b_0$-values of Thule and the Borysthenes.

Eratosthenes' $b_0$ of Thule corresponds to $46,400\,\text{st}/700\,\frac{\text{st}}{\degree}\approx66\degree17'$, which equals the actual position of the polar circle at Eratosthenes' time.
Apparently, he had a good knowledge of the value of $\varepsilon$, which he used in conjunction with Pytheas' information for the localisation of Thule.
Assuming for $b_0=46,400$\,st a resolution of 100\,st, the limits for $\varepsilon$ are:
\begin{align}
90\degree-(46,450/700)\degree &< \varepsilon < 90\degree-(46,350/700)\degree \\
23\degree39' &< \varepsilon < 23\degree47' \; .
\end{align}

Ptolemy states in his \textit{Mathematike Syntaxis} (MS; see \cite{man12},\linebreak
\cite{too84}) I.12 that the ratio $t=\frac{11}{83}$ of the arc between the tropics to the full meridian equals nearly Eratosthenes' value, which was also used by Hipparchus.
Ptolemy's $t$ leads to
\begin{equation}
\varepsilon_\text{m} = 23\degree51'20'' \; .\footnote{\cite[vol.\ 1, p.\ 44, fn.\ b]{man12} wrongly infers from MS~I.12 that $\varepsilon_\text{m}$ was Eratosthenes' value. According to the text, however, this applies only approximately (see also \cite[p.\ 459]{jon11}).}
\end{equation}
Hipparchus presumably used
\begin{equation} \label{eqn:epsh}
\varepsilon_\text{h}=23\degree40'
\end{equation}
(cf.\ \cite{dil34}).
Probably, this is Eratosthenes' value.
It corresponds to $t\approx\frac{10.91}{83}$, which does not differ significantly from Ptolemy's value.
For the polar circle, i.e.\ Thule, it yields $\phi=66\degree20'$ and $b_0=46,433\,\text{st}\approx46,400\,\text{st}$ in agreement with the value resulting from Eratosthenes' $b$-data.
Possibly Eratosthenes specified $b$ of Borysthenes -- Thule as ``about'' 11,500\,st because it was calculated from $b_0=46,433$ of Thule and $b_0=34,900$\,st of the Borysthenes so that 11,533\,st was obtained.
Or he considered the derived $b$ of Borysthenes -- Thule to be unreliable because $b_0$ of the Borysthenes was based on the lengths of the sea routes reported by navigators (cf. Section \ref{sec:searoutes}).

From F34 (G~II.5.7) can be derived that Eratosthenes' value for $\varepsilon$ was $\frac{4}{60}$ of $C$ (i.e.\ 24\degree).
This value is not contradictory to $23\degree40'$ because it is based on the division of the full circle into 60 parts.
It was a common value for $\varepsilon$ in early Greek geography, and Eratosthenes probably gave this rough value as well as a more precise value in his works.
Later ancient authors also mention or use this value, although a more accurate value was known (e.g. Ptolemy in GH~VII.6.7; see \cite[p.\ 734]{neu75}).
Strabo was probably not interested in Eratosthenes' more precise value so that he adopted the value  $\frac{4}{60}C$ only.
Similarly, in G~II.5.43 Strabo refers the reader to Hipparchus' work concerning astronomical matters.

%========================================================================
\section{Hipparchus' latitudinal data} \label{sec:hipp}
%========================================================================

The investigations of Hipparchus' latitudinal data are mainly based on the fragments (F) given by \cite{dic60} (the following translations are taken from it).
The data mainly originate from Hipparchus' treatise ``Against the `Geography' of Eratosthenes'', which consisted of three books (cf.\ \cite[p.\ 37]{dic60}).
Latitudinal data occurred in the second book (F12--34) and the third book (F35--63), the majority in the third book.
The third book contained astronomical data for several latitudes.
Strabo gives extracts of this compilation; for instance he limits the data to the inhabited world (cf.\ G~II.5.34).
The occurring types of latitudinal data are:
meridian arc length $b$ or $b_0$ between the parallels of two locations or from the equator (in st);
noon altitude $a$ of the sun at the winter solstice given in astronomical cubits (c; $1\,\text{c}=2\degree$); ratio $r=g:s$ of the length $g$ of the gnomon to the length $s$ of its shadow;
length $M$ of the longest day (summer solstice) in (equinoctial) hours.
Presumably, the meridian arc lengths do not originate from Hipparchus but were calculated by Strabo from latitudes by means of relation (\ref{eqn:degree_erat}) (e.g., \cite[p.\ 37]{ber69}).
The $M$-data are compared with their corresponding meridian arc lengths by \cite{raw09a}; for the sake of completeness, however, they are included in the following investigation.

With the exception of F15 (G~II.1.12),
the latitudinal data in the fragments of Hipparchus' second book (F19/G~II.1.12, F22/G~II.1.29, F24/G~II.1.34, F26/G~II.1.36) do not have connections to the data in Hipparchus' third book.
The only data which positions the concerned locations absolutely in latitude is the imprecise information that $b$ of Athens -- Babylon ``\ldots\ is not greater than 2,400 stades \ldots'' (F22).
Hipparchus gives this limit only in order to show that Eratosthenes' positioning of the Taurus is wrong.
Furthermore, there are no connections among the data of the second book which would cause redundant relations among each other.
Thus, these data are not included in the following investigations; for a discussion of the data see \cite{dic60}.

In places Strabo gives one latitudinal data which applies to several locations.
Among these locations there may be additions by Strabon which do not originate from Hipparchus (see \cite[p.\ 41]{ber69}).
Owing to uncertainties in this regard, however, all locations are taken into account here.
If within a fragment more than two locations are related by one data, the derivable relations are kept as compact as possible\footnote{In the case of information such as ``A, B, C are $x$\,st distant from D, E, F'', not all nine derivable distances from A, B, C to D, E, F are introduced but only the following five distances: A -- D: $b = x$\,st; A -- B, A -- C, D -- E, D -- F: $b = 0$.
This is advisable because the value $x$ was surely not determined for all nine distances.}.
The data which occur repeatedly within one fragment are listed and used once only.

A further source is Hipparchus' ``Commentary on the Phenomena of Aratus and Eudoxus'' (CP; see \cite{man94}), which contains only a few latitudinal data.
In the ``Commentary'' altitudes of the pole $a_\text{p}$ are given, which equal the latitude $\phi$, as well as polar distances $\zeta_\text{a}$ of the ever visible circle (the arctic circle) or of the never visible circle (which delimits the region of the stars which do not rise).

The considered data and their sources are composed in Table \ref{tab:hippdata}.\footnote{I thank one of the referees for the information that the value $b=12,500$\,st in G~II.1.18, which is given by \cite[p.\ 1313]{neu75} for Massalia -- 19\,h-parallel, is not from Hipparchus but from Strabo.}
The fragment numbers 15--61 correspond to \cite{dic60} and are extended by a consecutive number separated by a dot.
F62--71 refer to the ``Commentary'' and are additionally introduced here (only partly mentioned by \cite{dic60}, \cite[p.\ 54, F V 11]{ber69}, \cite{shc07a}).
Figure \ref{fig:hipp} shows the data in form of a graph.

For a comparison and a joint analysis of the consistency of the data, data not given as $b$ were converted into $b$.
The conversions of the given quantities into $\phi$ are considered in the following; $\phi$ can be converted into $b_0$ by means of (\ref{eqn:degree_erat}).
If further parameters are included in a conversion, it needs be considered for their choice whether the quantities to be compared were originally independently determined or not.

\begin{figure}[p]
\centering
\includegraphics[width=8cm]{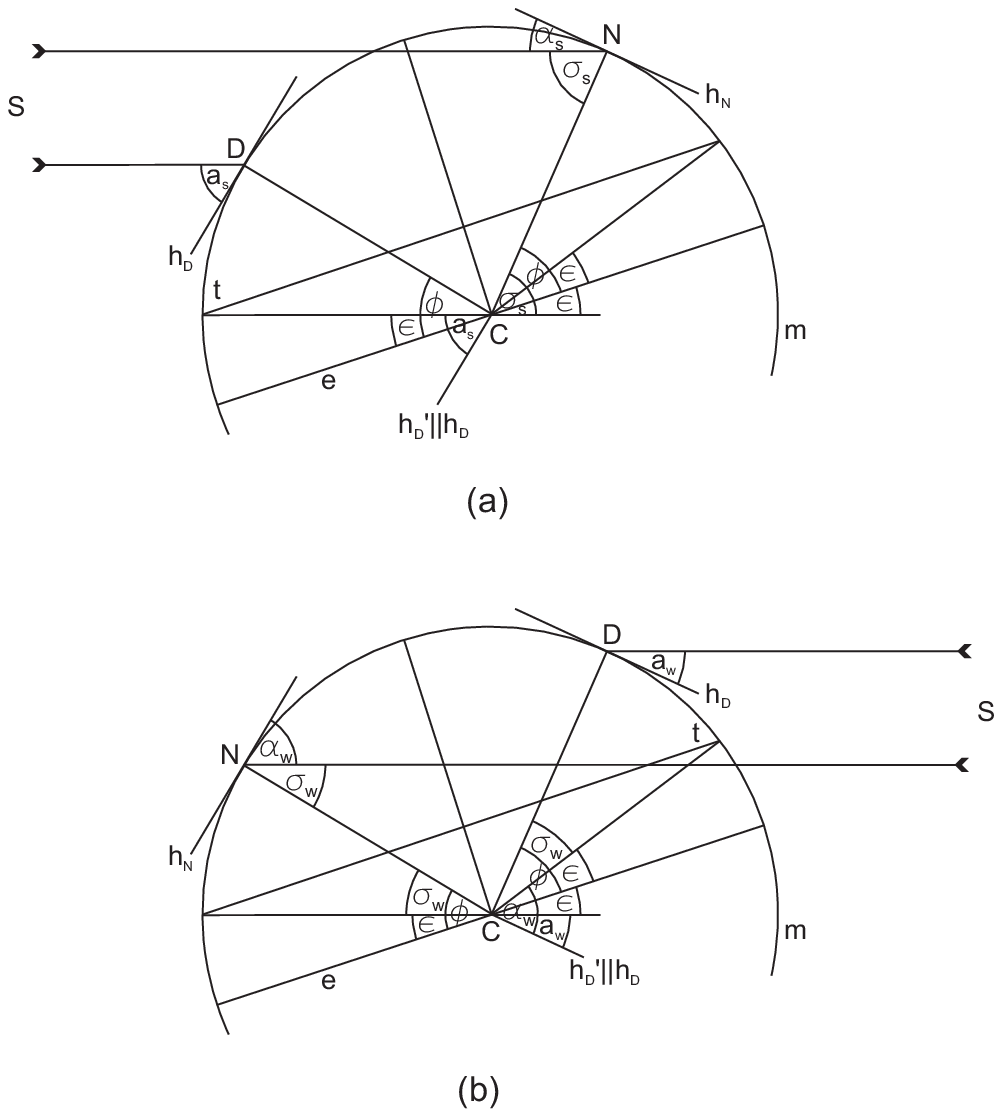}
\caption{On Strabo's information on Lake Maeotis in fragment F57, (a) summer solstice, (b) winter solstice (e: equator, h$_\text{D/N}$: horizon at noon/midnight, m: meridian, S: sun, t: tropic)}
\label{fig:maeotis}
\end{figure}

The conversion of $a$ into $\phi$ is
\begin{equation} \label{eqn:a}
\phi = 90\degree - a + \delta \; ,
\end{equation}
where $\delta$ is the declination of the sun.\footnote{\cite[p.\ 304]{neu75} states that the $a$-data form an arithmetical progression of the second order.
This is based, among others, on the value $a=3$\,c for $M=19$\,h.
In F61 (G~II.1.18), however, Strabo says that the $a$ belonging to $M=19$\,h is less than 3\,c.
An arithmetical progression is not considered here.}
If the sun altitude $a_\text{s}$/$a_\text{w}$ refers to the summer/winter solstice, $\delta$ equals
$\varepsilon$/$-\varepsilon$:
\begin{align}
\label{eqn:as}
\phi &= 90\degree - a_\text{s} + \varepsilon \\
\label{eqn:aw}
\phi &= 90\degree - a_\text{w} - \varepsilon
\end{align}
(for the case of $a_\text{s}$/$a_\text{w}$ see Figure \ref{fig:maeotis}(a)/(b)).
Since an ancient conversion from $\phi$ to $a$ is assumed here, for $\varepsilon$ the value of $\varepsilon_\text{h}$ (see (\ref{eqn:epsh})) is used which presumably underlies Hipparchus' conversion from $M$ to $\phi$ (see \cite{dil34}).

The ratio $r$ refers to the equinox or the summer solstice.
In the case of the equinox, $\phi$ is computed from ratio $r_\text{e}$ by
\begin{align} \label{eqn:phi_re}
\phi &= \arctan{(1/r_\text{e})} \; .
\end{align}
In the case of an equinoctial ratio, it is to be expected that it is not the result of a measurement because at the equinox only unreliable gnomon measurements are possible in contrast to the solstices (cf.\ \cite{raw09a}).
In the case of a ratio $r_\text{s}$ referring to the summer solstice, a real measurement can be expected and it holds true that
\begin{align}
\label{eqn:phi_rs}
\phi &= \arctan{(1/r_\text{s})} + \varepsilon \; .
\end{align}
In order to compare $r_\text{s}$ with an independently determined meridian arc length, for $\varepsilon$ the actual value must be used.
$\varepsilon=23\degree43'$ of Hipparchus' time is used here.
Furthermore, since the shadow is generated by the upper edge of the sun and not by its centre, $\phi(r_\text{s})$ must be enlarged by a systematic error of 16$'$, whereby the radius of the sun disc is taken into account (cf., e.g., \cite[p.\ 178]{dic60}).

Ptolemy gives the calculation of $\phi$ from $M$ and vice versa by means of spherical trigonometry in MS~II.3.
The comparison of Hipparchus' $M$-data with the associated $b_0$-data by \cite{dil34} and \cite{raw09a} suggests that Hipparchus
used a conversion $\phi_\text{t}(M,\varepsilon)$ based on spherical trigonometry too.
The modern formulation of Ptolemy's computation of $\phi$ from $M$ is
\begin{equation} \label{eqn:phi_M}
\phi_\text{t}(M,\varepsilon) = \arctan {( - \cos{( M/2 \cdot 15 \frac{\degree}{\text{h}} )} / \tan{\varepsilon} )}
\end{equation}
($M$ in h), which is applied here.
For $\varepsilon$ the value $\varepsilon_\text{h}$ is used, which probably underlies Hipparchus' conversion between $M$ and $\phi$ (see \cite{dil34}, \cite{raw09a}).
This value has only an inconsiderably small difference to the actual value 23\degree43$'$ in Hipparchus' time.
Since Hipparchus presumably converted $M$ into $\phi$ and Strabo $\phi$ into $b_0$, a conversion according to \cite{raw09a} is used here: $\phi_\text{t}(M)$ is rounded to the nearest $\frac{1}{12}\degree$ and $b_0(\phi_\text{t})$ to the nearest 100\,st.
The latter rounding is not applied to the data of the ``Commentary''.

For the conversion of $\zeta_\text{a} = 90\degree - \delta_\text{a}$ into $\phi$, equation (\ref{eqn:arctic}) applies so that $\phi = \zeta_\text{a}$.

%------------------------------------------------------------------------
\subsection{Notes on the data} \label{sec:hipp_notes}
%------------------------------------------------------------------------

The data on the Borysthenes refer to its mouth, cf.\ Section \ref{sec:erat_notes}.
In F57 (G~II.5.42), however, Strabo discusses ``the regions in neighbourhood of the Borysthenes and the southern parts of Lake Maeotis''.
These regions are distinguished from the Borysthenes and referred to as ``Lake Maoetis'' here.

F15 (G~II.1.12):
From this passage it follows that Mero\"{e} and the southern headlands of India have the same latitude. According to G~II.1.20, however, this is objected by Hipparchus in his second book (cf.\ \cite[pp.\ 42, 97]{ber69}) so that the information is not considered here.

F43 (G~II.5.35):
According to this passage, the Cinnamon country is situated ``\ldots\ very nearly half-way between the equator and the summer tropic \ldots\,.''
Following \cite[p.\ 44]{ber69}, it is assumed here that this inaccurate localisation is not from Hipparchus.

F46.3 (G~II.5.36):
\cite{dil34} indicates that $b_0=11,800$\,st of Mero\"{e}, which results from F43 (G~II.5.35), is contradictory to $b_0\approx11,600$\,st resulting from the conversion of $M=13$\,h of the associated \textit{klima}\footnote{The term \textit{klima} denoted a latitudinal strip or a latitude which was assigned to a specific $M$-value; in this regard see, e.g., \cite{hon29}, \cite[pp.\ 154--164]{dic60}, \cite[pp.\ 725--727]{neu75}.} in F46 (G~II.5.36).
\cite{raw09a} points out the difference between the city Mero\"{e} and the Mero\"{e}-\textit{klima} and gives $b_0$ for the \textit{klima}.
According to that, for $b_{0}^{\text{Mk}}$ of the Mero\"{e}-\textit{klima} and $b_{0}^{\text{Ak}}$ of the Alexandria-\textit{klima} follows from F46: $b_{0}^{\text{Mk}} + (b_{0}^{\text{Mk}} - 1,800\,\text{st}) = b_{0}^{\text{Ak}}$,
$b_{0}^{\text{Mk}} = (b_{0}^{\text{Ak}} + 1,800\,\text{st})/2$.
By means of $b_0^\text{Ak}=21,400\,\text{st}$ (see Section \ref{sec:hipp_results}) it follows that $b_{0}^{\text{Mk}}=11,600\,\text{st}$.

F47.3 (G~II.5.36):
``In Syene [\ldots] the sun stands in the zenith at the summer solstice \ldots\,.''
Thus, $a_\text{s} = 90\degree$ can be derived.
From (\ref{eqn:as}) follows $\phi = \varepsilon$, for which the actual value $23\degree43'$ is applied, since a real observation is assumed at first.
Hence, $b_0=16,602$\,st.

F48.4 (G~II.5.38):
Hipparchus distinguishes between Alexandria and the region 400\,st south of it (Alexandria-\textit{klima}, see F48.1) where $M$ is $14$\,h (F48.3).
For $r_\text{e}$ of Alexandria specified by Strabo (F48.4), \cite[p.\ 511]{jon17} and \cite[p.\ 95]{dic60} give $\frac{5}{3}$, which is an alteration.
The correct value of the text is $\frac{5}{7}$ (cf.\ \cite[p.\ 336]{neu75}, \cite{raw09a}).
Since this value yields a totally wrong $\phi$, it is assumed (cf.\ \cite[p.\ 336]{neu75}) that it is an $m:M$-ratio of the length $m$ of the shortest day to the length $M$ of the longest day, which was a usual quantity for the specification of the latitude.
From $m=24\,\text{h}-M$ follows $M=24\,\text{h}/(m/M+1)=14\,\text{h}$, which is used here.
Thus, contradictory to the text, this value does not refer to Alexandria but to the Alexandria-\textit{klima} (likewise \cite{raw09a}).

F48.6 (G~II.5.38):
Strabo gives $r_\text{e}=\frac{11}{7}$ for Carthage, which yields the grossly erroneous $\phi=32\degree28'$ by means of (\ref{eqn:phi_re}) (real $\phi=36\degree51'$).
\cite{raw85,raw09a} assumes a similar error for Carthage as for the Alexandria-\textit{klima} (see F48.4).
According to this, the given $r_\text{e}=\frac{11}{7}$ would be an $M:m$-ratio, which corresponds to the common \textit{klima} of $M=14\frac{2}{3}$\,h (cf.\ \cite[p.\ 722]{neu75}).
This explanation is not followed here.
First, the ratio $\frac{5}{7}$ for the Alexandria-\textit{klima} is assumed to be an $m:M$-ratio, but the ratio $\frac{11}{7}$ for Carthage would be an $M:m$-ratio so that a further inconsistency in the text would have to be assumed.
Second, Ptolemy gives $\phi=32\degree40'$ and $M=14\frac{1}{5}$\,h for Carthage (GH~IV.3.7, VIII.14.5).
Since Ptolemy's data rather originate from Hipparchus' data than from Strabo's data,\footnote{For example, Ptolemy states (GH~I.4.2) that altitudes of the pole originating from Hipparchus were available to him.} Hipparchus' $\phi$ must be about $32\degree40'$, which is fulfilled by $r_\text{e}$.

F52.2 (G~II.5.41):
In F52 the ratio $r_\text{s}=120/41\frac{4}{5}$ is given for Byzantium.
According to F53 (G~I.4.4) Hipparchus found the same ratio in Byzantium as Pytheas in Massalia.
If $r_\text{s}$ is Hipparchus' ratio for Byzantium, a real measurement by Hipparchus is unlikely because the ratio yields an error of about 2\degree\ with respect to the actual $\phi$.
\cite{jon02}, for example, assumes a calculative origin for $r_\text{s}$ and recalculates $1/r_\text{s}$ from $M=15\frac{1}{4}$\,h (F52.1) by means of $\varepsilon_\text{h}$, $\varepsilon_\text{m}$ and $\varepsilon_{r}$ (cf.\ Sections \ref{sec:erat_notes}, \ref{sec:obliquity}) but does not achieve identicalness with $120/41\frac{4}{5}$.\footnote{In order to test the possibility of an ancient conversion from $M$ to $r_\text{s}$, the conversion of \cite{jon02} is redone with differing calculation steps in the following.
The original conversion from $M$ to $\phi$ was presumably based on $\varepsilon_\text{h}$ (\cite{dil34}).
Its result was probably rounded to the nearest $\frac{1}{12}\degree$: $\phi_\text{t}(M)=43\degree17'\approx43\degree15'$.
The conversion from $\phi$ to $r_\text{s}$ is based on (\ref{eqn:phi_rs}), i.e.\ on the determination of $\tan{(\phi-\varepsilon)} = \tan{\alpha}$.
As it can be expected from ancient calculations, the tangent-function was determined by the ratio $\operatorname{crd}(2\alpha)/\operatorname{crd}(180\degree-2\alpha)$ (following from $2\sin{\alpha}=\operatorname{crd}2\alpha$, see \cite[pp.\ 21--24]{neu75}).
The chord $\operatorname{crd}()$ is determined here by a linear interpolation of Hipparchus' presumable table of chords, which was reconstructed by \cite{too74}.
The results for $\varepsilon_\text{h}$, $\varepsilon_\text{m}$, $\varepsilon_\text{r}$ are $\approx42.69/120$, $\approx42.25/120$ and $\approx41.91/120$, respectively.
The resulting numerator, however, should be within the interval $[41\frac{3.5}{5}=41.7, 41\frac{4.5}{5}=41.9]$.}
Following \cite{raw09a}, it is assumed here that $r_\text{s}$ is the result of a real measurement which Pytheas performed in Massalia.
Hence, $\varepsilon=23\degree44'$ of Pytheas' time is used for conversion (\ref{eqn:phi_rs}).

F56 (G~II.5.41):
``If one sails into the Pontus [Black Sea] and proceeds about 1,400 stades [F56.1] northwards, the longest day becomes 15$\frac{1}{2}$ equinoctial hours [F56.2].''
The distance refers to the parallel of Byzantium (cf.\ \cite[p.\ 183]{dic60}).
Furthermore, the mentioned region in the Pontus (Mid-Pontus) is ``\ldots\ equidistant from the pole and the equator \ldots'' so that $b_0=C/8=31,500$\,st (cf.\ (\ref{eqn:C})) is used (F56.3).
Moreover, there ``\ldots\ the arctic circle is in the zenith \ldots'' (i.e.\ only one point of the circle).
Hence, its declination $\delta_\text{a}$ equals $\phi$ and (\ref{eqn:arctic}) yields $\phi=90\degree/2=45\degree$ or $b_0=C/8$, which is not introduced here once more.

F57 (G~II.5.42):
Strabo reports on the neighbourhood of the Borysthenes and the southern parts of Lake Maeotis: ``The northern part of the horizon, throughout almost the whole of the summer nights, is dimly illuminated by the sun [\ldots]; for the summer tropic is seven-twelfths of a zodiacal sign from the horizon [$=a_\text{t}$], and therefore this is also the distance that the sun is below the horizon at midnight [$=\alpha$].''
One zodiacal sign corresponds to $360\degree/12=30\degree$ and $\frac{7}{12}$ of a sign is 17\degree30$'$.
The text suggests that the angles $a_\text{t}$ and $\alpha$ refer to the summer solstice, that $a_\text{t}$ is 17\degree30$'$ and that $\alpha$ equals $a_\text{t}$.
Figure \ref{fig:maeotis} shows the meridian m, the equator e and the positions D and N of the observer at noon and at midnight, respectively, at the summer (a) and winter (b) solstice.
Shifting the horizon h$_\text{D}$ in the centre C of the earth to position h$'_\text{D}$, m can be regarded as the celestial sphere.
Then, the sun altitude $a_\text{s}$ at C in Figure (a) and the angle $\alpha_\text{w}$ at C in Figure (b) correspond to Strabo's description of $a_\text{t}$ and $\alpha_\text{s}$ at N in Figure (a) and $\alpha_\text{w}$ at N in Figure (b) correspond to Strabo's description of $\alpha$.
Owing to $\alpha_\text{s}\neq a_\text{s}$, Strabo's equation $\alpha = a_\text{t}$ holds true for the winter solstice only.
For the summer solstice, $a_\text{s}=90\degree-\phi+\varepsilon\approx64\degree57'$ (equation (\ref{eqn:as}) with $b_0=34,100\,\text{st}$ of Lake Maeotis, cf.\ Section \ref{sec:hipp_results}) applies, which is inconsistent with the given value 17\degree30$'$ of $a_\text{t}$.
The altitude $a_\text{w}=90\degree-\phi-\varepsilon\approx17\degree37'$ (equation (\ref{eqn:aw})) of the winter solstice, however, is consistent with the value of $a_\text{t}$ but inconsistent with Strabo's description of $a_\text{t}$.
From Figure (a) follows $\alpha_\text{s}=90\degree-\sigma_\text{s}=90\degree-\phi-\varepsilon=a_\text{w}$ (equation (\ref{eqn:aw})); from Figure (b) follows $\alpha_\text{w}=90\degree-\sigma_\text{w}=90\degree-\phi+\varepsilon=a_\text{s}$ (equation (\ref{eqn:as})).
In summary, Strabo's information can be corrected by the following two statements.
First, $a_\text{t}$ is (about) 65\degree\ at the winter solstice and equals $\alpha$ at the winter solstice ($=\alpha_\text{w}$).
Second, $a_\text{w}$ is 17\degree30$'$ and equals $\alpha$ at the summer solstice ($=\alpha_\text{s}$).
Nonetheless, Strabo's information is not used further.

F60.1 (G~II.5.42):
\cite[p.\ 28]{gos98}\footnote{I thank one of the referees for his reference to \cite{gos98} in the context of the errors of F60.1 and F61.3.}, \cite[p.\ 70]{ber69} and \cite{dil34} notice that $b=6,300$\,st for Byzantium -- ``north of Lake Maeotis'' is 1,400\,st (2\degree) too small in comparison to the $M$-data of these locations (F52.1: 15$\frac{1}{4}$\,h, F60.3: 17\,h).
\cite{dil34} assumes that Strabo inadvertently used Mid-Pontus (F56.2: $M=15\frac{1}{2}$\,h) instead of Byzantium for the calculation of $b$.
Accordingly, the corrected $b=7,700$\,st is used here.

F61.1 (G~II.1.18):
\cite[p.\ 185]{dic60} shows that $b=6,300$\,st of Massalia -- Celtica (according to Hipparchus, but according to Strabo north of Celtica) has also an error of 2\degree\ as $b$ of F60.1.
It is corrected to 7,700\,st here.

F61.3 (G~II.1.18):
$b=9,100$\,st of Massalia -- 18\,h-region has the same error of 2\degree\ as $b$ of F60.1 (\cite[p.\ 28]{gos98}, \cite[p.\ 70]{ber69}, \cite{dil34}).
The corrected $b=10,500$\,st is applied here.

F63 (CP~I.3.7):
Hipparchus refers to regions at the Hellespont.
These regions are equated here with Alexandria in the Troad.
Hipparchus gives $M:m=\frac{5}{3}$ and $M=15$\,h.
Since both are equivalent, only the latter is used here.

F65.1 (CP~I.4.8):
Hipparchus says that $\zeta_\text{a}$ is about 37\degree\ in the environment of Athens and there where $r_\text{e}=\frac{4}{3}$.
Although he does not explicitly assign this $r_\text{e}$ to Athens, it is used here for Athens.

F66.1, F67.1, F71.1 (CP~I.7.11, I.7.14, II.4.2):
Hipparchus gives the same data of the culmination, rising and setting of constellations for Greece as for the regions where $M=14\frac{1}{2}$\,h.
Thus, this $M$-value is assigned to Greece here.

%------------------------------------------------------------------------
\subsection{Test of consistency} \label{sec:hipp_results}
%------------------------------------------------------------------------

\begin{figure}[p]
\centering
\includegraphics[width=13cm]{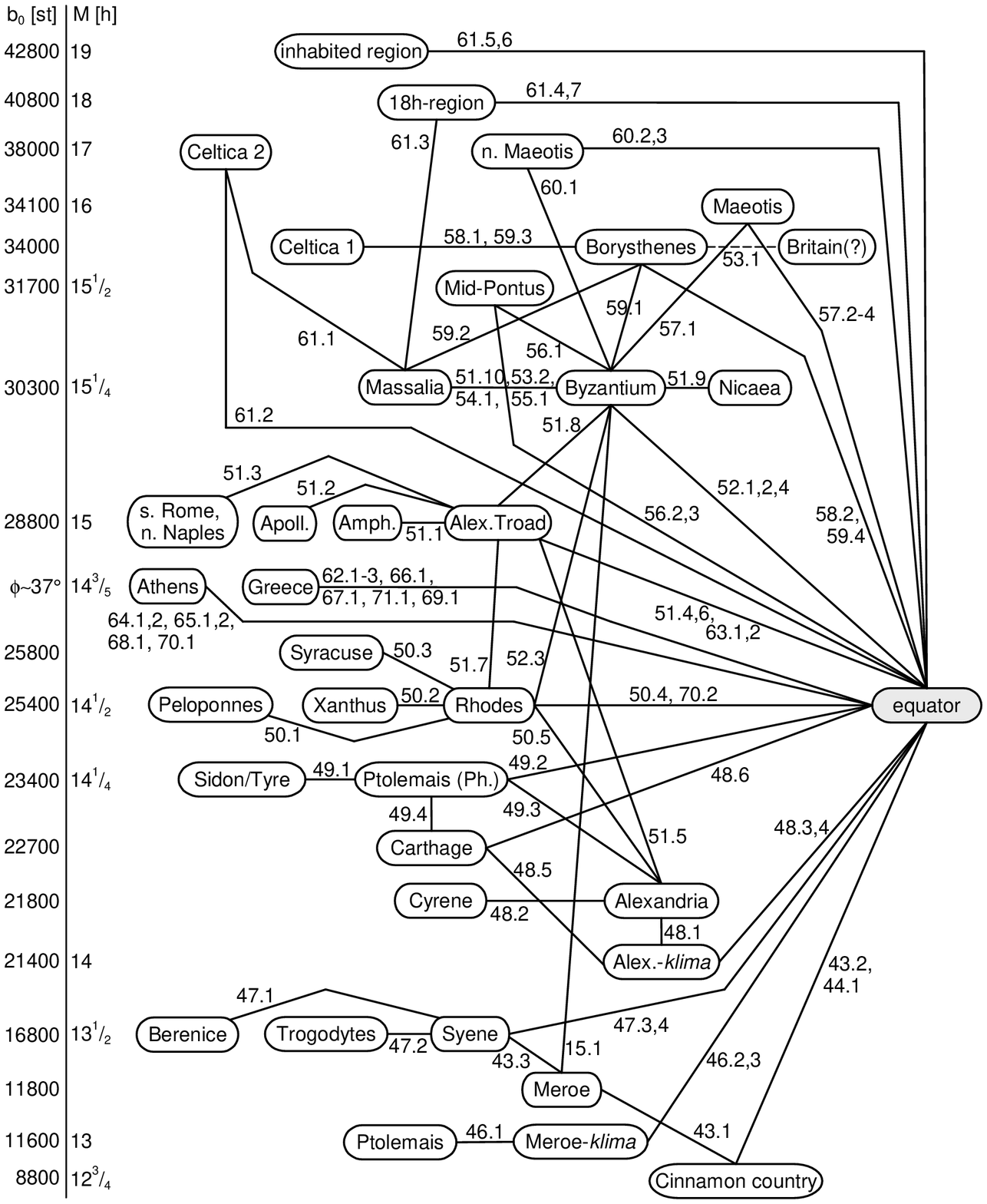}
\caption{Graph of the latitudinal data ascribed to Hipparchus; the vertical order of the locations gives the meridian arc length $b_0$ from the equator (not drawn to scale; s. = south of, n. = north of)}
\label{fig:hipp}
\end{figure}

The consistency of the data ascribed to Hipparchus is tested according to Section \ref{sec:erat_results}.
The $n=84$ data $b_i$ given in Table \ref{tab:hippdata} are composed to the equation system (\ref{eqn:obs}).
The imprecise $a_\text{w}$ of F61.5 is not involved in the adjustment computation.
Furthermore, the $M$-data of F66.1, F67.1 and F71.1 are not used because probably they are imprecise values (see below).
There remain $n=80$ data for the adjustment computation.
The equation system (\ref{eqn:obs}) has 34 unknown $b_0$ of the locations shown in Figure \ref{fig:hipp} (for Celtica only one $b_0$ is used).
The $L_1$-norm adjustment by means of the BR-algorithm yields 61 $v_i$ being 0; hence, the related data are consistent among each other.
The $v_i\neq0$ are considered in the following.
It turns out that the solution for the $b_0$ can be regarded as to be in accord with Hipparchus' original data.
The $b_0$ are given in Figure \ref{fig:hipp}.

F15.1; Mero\"{e} -- Byzantium; $b\approx18,000$\,st; $v=500$\,st:
18,000\,st are contradictory to $b=18,500$\,st which follows from $b_0$ of F43.2 (Cinnamon country), $b_0$ of F52.3 (Byzantium) and $b$ of F43.2/44.1 (Cinnamon country -- Mero\"{e}).
Since, however, Strabo gives ``about'' 18,000\,st, there is not a real contradiction.

F47.3; Syene; sun at zenith at summer solstice $\Rightarrow b_0=16,602\,\text{st}$; $v=198$\,st:
$b_0$ was derived from $\phi=23\degree43'=\varepsilon$ (Section \ref{sec:hipp_notes}); it is contradictory to $b_0=16,800$\,st which follows from $b_0$ of F43.2 (Cinnamon country), $b$ of F43.1 (Cinnamon country -- Mero\"{e}) and $b$ of F43.3 (Mero\"{e} -- Syene).
16800\,st corresponds to $\phi=24\degree$ in good agreement with the real $\phi=24\degree05'$.
Thereby, it equals the common ancient value $\varepsilon_\text{r}$ so that Syene was theoretically located on the tropic.
Nonetheless, the information of F47.3 may be based on a real observation.
Owing to the closeness to the tropic, the sun altitude has been about 90\degree\ at noon at the summer solstice so that for an observer the sun apparently stood in the zenith.\footnote{For 220--120 BC, the maximal sun altitude at the summer solstice was determined based on the calculation method given in Fn.\ \ref{foo:altitude}; the result is $89\degree37'\approx90\degree$.}

F48.6; Carthage; $r_\text{e}=\frac{11}{7} \Rightarrow b_0=22,730\,\text{st}$; $v=-30$\,st:
$b_0$ of F48.3 (Alexandria-\textit{klima}) and $b$ of F48.5 (Alexandria-\textit{klima} -- Carthage) as well as $M$ of F49.2 (Ptolemais in Phoenicia) and $b$ of F49.4 (Carthage -- Ptolemais) yield $b_0=22,700$\,st for Carthage.
If this value was calculated from $\phi$ and rounded to the nearest 100\,st, the original $b_0$ should be within the interval $[22,650\,\text{st}, 22,750\,\text{st}]$.
This is fulfilled by the $b_0$ derived from $r_\text{e}$.
Hence, $r_\text{e}$ could have been calculated from $\phi$.

F50.5; Alexandria -- Rhodes (centre); $b=3,640$\,st; $v=-40$\,st:
$b$ is contradictory to the value 3,600\,st which follows from $b=7,000$\,st of F51.5 (Alexandria -- Alexandria in the Troad) and $b=3,400$\,st of F51.7 (Rhodes -- Alexandria in the Troad).
\cite{dil34} assumes an error of 40\,st in $b=3,640$\,st originating from a faulty reading.
\cite[p.\ 53]{ber69} states that the value of 3,640\,st is given with a higher precision and that it refers to the city Rhodes.
\cite[p.\ 176]{dic60} considers $b$ to be derived from a real measurement by Hipparchus in the centre of Rhodes and assumes that the text originally gave 3440\,st.
\cite{shc07a} assumes that both 3,600\,st and 3,640\,st are authentic and that they refer to the centre of Rhodes and the city Rhodes, respectively.
The following explanation shall be added.
In F50 Strabo refers not only to the centre of Rhodes but to ``\ldots the regions round the centre of Rhodes \ldots'' and states that there $M$ is $14\frac{1}{2}$\,h.
Hipparchus assigned a $\phi$-value to the $14\frac{1}{2}$\,h-\textit{klima}, which probably was $\phi_1=\phi_\text{t}(M) = 36\degree15'$.
Strabo converted $\phi_1$ into $b_0$ and rounded it to the nearest 100\,st: $b_{01}(\phi_1)=25,375\,\text{st} \approx 25,400\,\text{st}$.
Strabo had a further value $\phi_2$ for the city Rhodes in the north of the island (e.g., from Hipparchus, who lived on the island Rhodes); it corresponded to $b_{02}=25,440$\,st ($=3640\,\text{st}+21800\,\text{st}$ of Alexandria), i.e.\ $\phi_2\approx36\degree21'$.
(That value is somewhat less than the real $\phi=36\degree26'$ and therefore consistent with the ancient systematic error of a gnomon measurement due to the generation of the shadow by the upper edge of the sun, see Section \ref{sec:hipp}.)
Strabo knew that $M\approx14\frac{1}{2}$\,h is spaciously valid, that $\phi_1$ is a theoretical value derived from $M$ and that his rounding up of $b_{01}$ yielded a less accurate and more northerly position (as $b_{02}$).
Furthermore, $b_{01}$ and $b_{02}$ only differ by 40\,st.
Thus, Strabo chose the more precise and trustable $b_{02}$ for his statement in F50 on the $14\frac{1}{2}$\,h-\textit{klima}.

F52.2; Byzantium; $r_\text{s}=120/41\frac{4}{5} \Rightarrow b_0=30,243\,\text{st}$; $v=57$\,st:
$r_\text{s}$ is assumed to be the result of an independent measurement (cf. Section \ref{sec:hipp_notes}); nevertheless, the small $v$ shows that $r_\text{s}$ is in accord with the other data.

F56.3; Mid-Pontus; $b_0=C/8 \Rightarrow b_0=31,500\,\text{st}$; $v=200$\,st:
$b$ is contradictory to $b_0=31700$\,st which follows, for instance, from $M$ of Mid-Pontus (F56.2).
However, $v$ is acceptable because $b_0$ of F56.3 is derived from rough information (cf.\ Section \ref{sec:hipp_notes}).

F57.4; Lake Maeotis; $a_\text{w}=9\,\text{c} \Rightarrow b_0=33,833\,\text{st}$; $v=267$\,st:
Since $a_\text{w}$ is a rounded value, it can be regarded as consistent with the other data if $v$ is $<0.5\,\text{c}=1\degree\,\hat{=}\,700\,\text{st}$.
That is fulfilled.

F58.2; Borysthenes; $a_\text{w}=9\,\text{c} \Rightarrow b_0=33,833\,\text{st}$; $v=167$\,st:
Cf.\ F57.4.

F60.2; ``north of Lake Maeotis''; $a_\text{w}=6\,\text{c} \Rightarrow b_0=38,033\,\text{st}$; $v=-33$\,st:
Cf.\ F57.4.
$a_\text{w}$ is consistent with $b_0=38,000$\,st which follows from $b_0$ of F52.4 (Byzantium) and the corrected $b=7,700$\,st of F60.1 (Byzantium -- ``north of Lake Maeotis'').

F61.1; Massalia -- Celtica; $b=7,700$\,st; $v=-4,000$\,st:
$b$ is contradictory to the value 3,700\,st which follows from $b$ of F59.2 (Massalia -- Borysthenes) and $b=0$ of F58.1/59.3 (Borysthenes -- Celtica).
From F61.1 and $b_0=30,300$\,st of Massalia follows $b_0=38,000$\,st for Celtica in contrast to $b_0=34,000$\,st which follows from F59.2 and F58.1/59.3.
This is not a real contradiction because Celtica is a region with a large latitudinal extent and Hipparchus did not distinguish between the Celtic and the Germanic coast (see \cite[pp.\ 185, 188]{dic60}) so that he gave a southern (F58.1, F59.3) and a northern (F61.1, F61.2) latitude for Celtica.
This becomes evident by Strabo's statement (F61/G~II.1.18) that Hipparchus takes the inhabitants of the region concerning F61.1 ``\ldots\ to be still Celts \ldots'' and that Strabo himself considers them as ``\ldots\ Britons who live 2,500 stades north of Celtica \ldots'' (``Celtica'' refers to Hipparchus' southern latitude).
Strabo's $b=2,500$\,st must be corrected by $+2\degree\mathrel{\widehat{=}}1,400\,\text{st}$ as $b$ of F61.1 (see Section \ref{sec:hipp_notes}).
Then, for the northern latitude of Celtica $b_0=(34,000+2,500+1,400)\,\text{st}=37,900\,\text{st}$ is obtained in accordance with 38,000\,st derived from F61.1.

F61.2; Celtica; $a_\text{w}=6\,\text{c} \Rightarrow b_0=38,033\,\text{st}$; $v=-4,033$\,st:
As F61.1 (Massalia -- Celtica), $a_\text{w}$ refers to the northern latitude of Celtica at $b_0=38,000$\,st.
The $v$ in this regard is only $-$33\,st, which is acceptable, cf.\ F57.4.

F61.4; 18\,h-region; $a_\text{w}=4\,\text{c} \Rightarrow b_0=40,833\,\text{st}$; $v=-33$\,st:
Cf.\ F57.4.

F61.5; ``inhabited region''; $a_\text{w}<3\,\text{c} \Rightarrow b_0>42,198\,\text{st}$; $v=567$\,st:
The information $b_0>42,198$\,st is in accord with $v>0$ and $v$ is acceptable because of the imprecise data; therefore, F61.5 is consistent.

F62.1, F65.1; Greece, Athens; $r_\text{e}=\frac{4}{3} \Rightarrow b_0=25,809\,\text{st}$; $v=91$\,st:
From $r_\text{e}$ and (\ref{eqn:phi_re}) follows $\phi=36\degree52'$.
Thus, $r_\text{e}$ is consistent because it is in accord with the latitude of 37\degree\ which follows from F62.3, F69.1 (Greece) as well as from F64.2, F65.2, F68.1, F70.1 (Athens).

F62.2, F64.1; Greece, Athens; $M=14\frac{3}{5}\,\text{h} \Rightarrow b_0=26,024\,\text{st}$; $v=-124$\,st:
$\phi_\text{t}(M,\varepsilon_\text{h})=37\degree18'$.
The difference to 37\degree\ of F62.3, F69.1 (Greece) as well as of F64.2, F65.2/70.1, F68.1 (Athens) is 18$'$.
\cite{shc07a} considers 37\degree\ as inconsistent with $\varepsilon_\text{h}$.
$\phi_\text{t}(M,\varepsilon=23\degree51')=37\degree03'$ is in better agreement so that \cite[p.\ 167]{dic60} assumes 23\degree51$'$ (MS~I.12) to be Hipparchus' value for $\varepsilon$.
However, $\varepsilon_\text{h}$ need not be refused.
First, Hipparchus only gives ``about 37\degree'' in F62.3 (Greece) as well as in F64.2, F65.2, F70.1 (Athens).
Second, Hipparchus usually uses a step width of $\frac{1}{4}$\,h or a multiple of it for his \textit{klimata}; the nearest $M$-values 14$\frac{1}{2}$\,h and 14$\frac{3}{4}$\,h yield 36\degree15$'$ and 38\degree47$'$ so that 14$\frac{3}{5}$\,h represents a good fit with 37\degree\ and can be regarded as consistent.
Moreover, Hipparchus assigns 14$\frac{3}{4}$\,h (F62.2) as well as 14$\frac{1}{2}$\,h (e.g., F66.1) to Greece, which illustrates Hipparchus' low demand for the accuracy of the $M$-data.

F63.1; Alexandria in the Troad;	$M=15\,\text{h} \Rightarrow b_0=28,753\,\text{st}$; $v=47$\,st:
$M$ is consistent, it is in agreement with F51.4.

F63.2; Alexandria in the Troad; $\phi=a_\text{p}\approx41\degree \Rightarrow b_0 \approx 28,700\,\text{st}$; $v=100$\,st:
$v$ is acceptable because of the approximate $a_\text{p}$ so that F63.2 is consistent.

F66.1, F67.1, F71.1; Greece; $M=14\frac{1}{2}\,\text{h} \Rightarrow b_0=25,308\,\text{st}$; $v=592$\,st:
$M$ is contradictory to $M=14\frac{3}{5}$\,h of F62.2.
The smaller $M=14\frac{1}{2}$\,h leads to a region south of Athens because Hipparchus assigns Athens to $M=14\frac{3}{5}$\,h (F64.1).
Hipparchus probably only gives a less accurate $M$-value with a resolution of $\frac{1}{2}$\,h in F66.1, F67.1, F71.1.

F70.2; Rhodes; $\zeta_\text{a} = 36\degree \Rightarrow b_0=25,200\,\text{st}$; $v=200$\,st:
From $M=14\frac{1}{2}$\,h of F50.4 follows $\phi_\text{t}(M)=36\degree15'$.
Hence, $\zeta_\text{a}$ of F70.2 is probably only a rough value as $a_\text{p}$ of Athens of F64.2.

Since the $a_\text{w}$-data of F60.2, F61.2 and F61.4 are inconsistent with the uncorrected textual $b$-values of F60.1, F61.1 and F61.3 (see Section \ref{sec:hipp_notes}), they confirm that the error of 1400\,st of these $b$-values is caused by Strabo.

In F51 Strabo assigns Alexandria in the Troad to the parallel which has $M=15$\,h (F51.4) and is ``over $28,800$\,st'' from the equator (F51.6).
From $M$ follows $\phi_\text{t}(M)\approx40\frac{1}{6}\degree\mathrel{\widehat{=}}28,817\,\text{st}\approx28,800\,\text{st}$ (which corresponds to the result of the adjustment).
In his statement Strabo possibly refers to the value of 28,817\,st, which resulted from his conversion of $\phi$ into $b_0$.

According to F53 the parallel through the mouth of the Borysthenes runs through Britain too (F53.1).
It is likely that Hipparchus referred to Celtica and Strabo replaced it by Britain (likewise \cite[p.\ 66, fn.\ 1]{ber69}) for the following reasons.
First, according to F58.1/59.3 (G~II.1.18/12) Hipparchus locates Celtica and the Borysthenes at the same latitude.
Second, according to F61 (G~II.1.18) Hipparchus locates Britain north of the ``inhabited region''/19\,h-parallel, i.e.\ much more northernly than the Borysthenes.
Last, according to F61 Strabo believes the Celts mentioned by Hipparchus to be Britons.

\cite[p.\ 184]{dic60} assumes that Strabo's data on the Borysthenes including F57 always refer to its mouth.
There is, however, evidence that Hipparchus located the mouth of the Borysthenes further south than the neighbouring region of the 16\,h-\textit{klima} (F57).
Strabo explicitly states that Hipparchus locates the mouth 3,700\,st north of Massalia and Byzantium (F59.1, F59.2) and 34,000\,st north of the equator (F59.4); from the former value also follows $b_0=34,000$\,st.
Furthermore Strabo says that $M$ is 16\,h in the regions in neighbourhood of the Borysthenes and the southern parts of Lake Maeotis (F57.2), which are 3,800\,st north of Byzantium (F57.1) and 34,100\,st north of the equator (F57.3); from the former $b$-value also follows $b_0=34,100$\,st.
The $b$-data are confirmed by $\phi_\text{t}(M)\approx48\frac{9}{12}\degree \mathrel{\widehat{=}} 34125\,\text{st}\approx34,100\,\text{st}$.
Hence, Hipparchus distinguished between the mouth of the Borysthenes and the 16\,h-\textit{klima}, which is 100\,st further north.

%========================================================================
\section{Summary}
%========================================================================

The latitudinal data attributed to Eratosthenes and Hipparchus were each compiled and formulated as systems of equations, whose solution revealed the differences and inconsistencies of the data.
As a result, the presumably original data of Eratosthenes and Hipparchus were deduced.

The analysis of the data ascribed to Eratosthenes showed several disagreements, which suggests that the concerned data originate from Strabo and not from Eratosthenes; this applies to F34.9, F34.10, F34.11, F35.5, F35.6, F36.2 and F47.1.
In particular, Eratosthenes' latitudinal extent of the inhabited world up to the ``northern regions'' (F30.1, F34.12) is contradictory to the corresponding sum of the given meridian arc lengths ascribed to Eratosthenes so far.
Therefore, Eratosthenes' meridian arc length of the part Borysthenes -- ``northern regions'' is probably not 4,000\,st (F34.9) but 3,000\,st, which is given by Strabo in G~II.5.8 (F34.13).

Eratosthenes' latitudinal distances Alexandria -- Hellespont -- Borysthenes (F35.2, F35.3/36.3) are grossly erroneous.
According to Strabo it was generally agreed that the sea route Alexandria -- Borysthenes is a straight line.
Hence, Eratosthenes presumably based his latitudinal distances Rhodes -- Hellespont -- Byzantium on the lengths of sea routes, which is affirmed by a good agreement of his distances with the actual distances alongside the Turkish coast.

From Pytheas' information on the position of the arctic circle relating to Thule it was known that Thule is situated at a latitude of $(90\degree-\varepsilon)$, where $\varepsilon$ is the obliquity of the ecliptic.
In conjunction with Eratosthenes' latitudinal data for Thule, 23\degree40$'$ can be derived for Eratosthenes' value of $\varepsilon$.
This value corresponds to Hipparchus' presumable value (see \cite{dil34}) and was possibly referred to by Ptolemy in MS~I.12.

The fragments ascribed to Hipparchus contain latitudinal quantities of different types.
Occurring differences of the data were explained by the different types of information and their different precision and origination.
The real inconsistencies can be ascribed to Strabo in most cases; this applies to F48.4 (e.g., \cite{neu75}), F60.1 and F61.3 (\cite{dil34}), F61.1 (\cite{dic60}), F15.1, F51.6, F53.1 and F57.
Strabo's statement on the distances of the summer tropic and the sun with respect to the horizon at Lake Maeotis in F57 has not been interpreted so far; his error in this regard was illustrated.
Hipparchus distinguished between the 14\,h-\textit{klima} and the city Alexandria as well as between the 13\,h-\textit{klima} and the city Mero\"{e} (\cite{raw09a}).
The present investigation revealed that Hipparchus probably also distinguished between the 16\,h-\textit{klima} and the mouth of the Borysthenes 100\,st south of the parallel of the \textit{klima}.

%========================================================================
\section*{Appendix: On the location of Thule} \label{sec:thule}
%========================================================================

Pytheas' voyage to Thule took place in ca.\ 330 BC.\footnote{\cite[p.\ 48]{nan11}: 330--325 BC; \cite[p.\ 162]{hen44}: 350--310 BC.}
His treatise ``On the Ocean'' on his voyages is not preserved, but later ancient authors provided extractions of this treatise.
The handed down information on Thule is given by, e.g., \cite[pp.\ 155--159]{hen44} and \cite{whi82}.
The main sources are Strabo's ``Geography'' and Pliny's ``Natural History''.
The only quotation from Pytheas' treatise is to be found in Geminus' \textit{Eisagoge} (E; see \cite{man98}).
Furthermore, Ptolemy describes the position and form of the island Thule by means of longitudes and latitudes in GH~II.3.

The two common localisations of Pytheas' Thule are Iceland (e.g., \cite{bur75}, \cite[p.\ 127]{rol10}) and Norway.
Iceland is neglected here because Pytheas met inhabitants in Thule according to E~VI.9, but so far a settlement of Iceland can not be assumed for his time.
\cite[p.\ 62]{nan11} and \cite[p.\ 166]{hen44} locate Thule in the region of Trondheim in Norway.

Ptolemy does not refer to Pytheas' Thule; his island Thule is usually identified as Shetland (e.g., \cite[p.\ 146]{riv79}).
Detailed reasons for this are given by \cite{mar13c} in an investigation of Ptolemy's coordinates of Scotland.
It should be added that furthermore Ptolemy's length of the longest day in Thule of 20\,h (GH~VIII.3.3, MS~II.6) contradicts Pytheas' information on the length of the nights in Thule (see below).

For a localisation of Pytheas' Thule, the following information comes into consideration:\\
1. Thule is a six-day seafaring from Britain in northern direction (G~I.4.2, NH~II.77).\\
2. In the region of Thule the tropic of Cancer coincides with the arctic circle (G~II.5.8).
At the summer solstice there are no nights (NH~IV.30).\\
3. The meridian arc length $b$ of Borysthenes -- Thule is 11,500\,st (G~I.4.2).\\
4. Pytheas said that in Thule the length $n$ of the nights was 2\,h and 3\,h (E~VI.9).\\
5. A one-day journey from Thule is the frozen/clotted sea (NH~IV.30).

The frozen/clotted sea suggests a larger appearance of sea ice.
That disagrees with the location of Thule in Norway because in the Norwegian Sea there is no drift ice (cf.\ \cite{vin91}) and at Pytheas' time, at the beginning of the Subatlantic, the climate was similar to today's climate so that drift ice can be excluded.
According to \cite[pp.\ 105, 156, fn.\ 1]{hen44} the clotted sea is a fiction, which can be found similarly in ancient and medieval literature.
Thus, information 5 does not play a role here.

Information 2 and 3 are treated in Section \ref{sec:obliquity}; information 1 and 4 are dealt with in the following.

\begin{figure}[p]
\centering
\includegraphics[width=8cm]{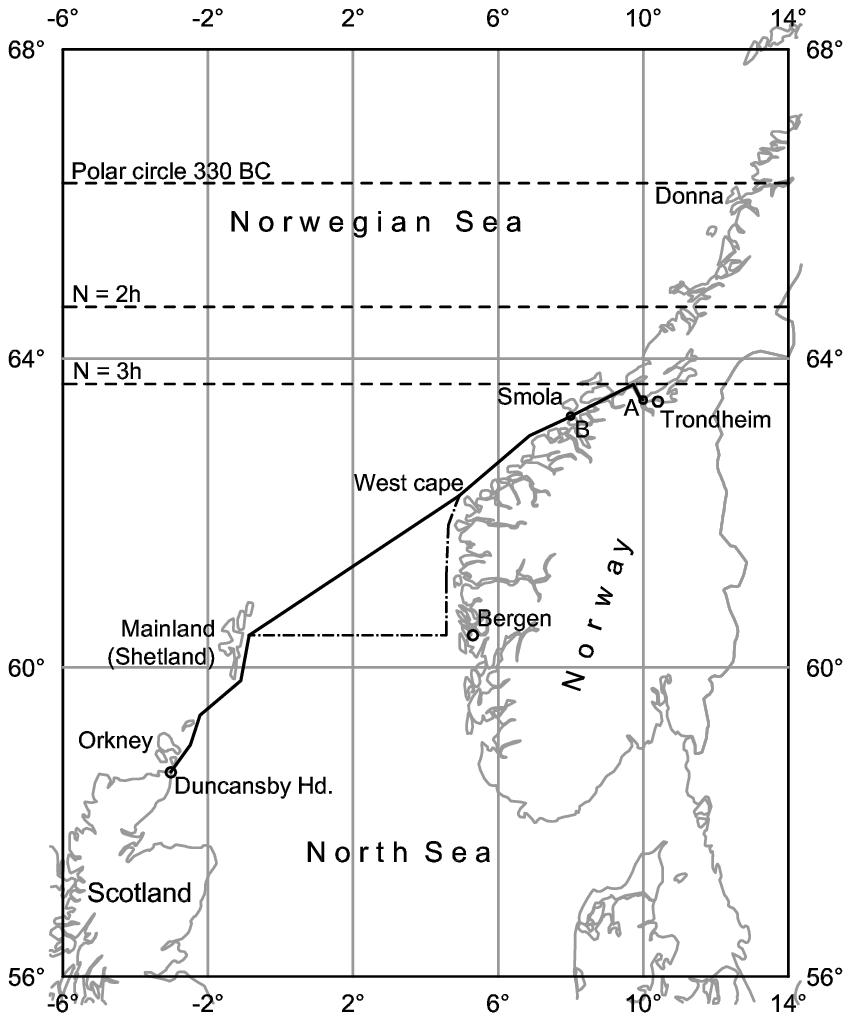}
\caption{Pytheas' possible sea route from Great Britain to Thule; route A: \textit{continuous}, route B: \textit{dash-dot}/\textit{continuous}}
\label{fig:reise}
\end{figure}

For the conversion of the time of six days (i.e.\ days and nights) into a length of a sea route, there is the following information.
First, Strabo says (G~X.4.5) that the sea voyage from Cyrenaea (Cape Rasat) to Criumetopon (Cape Krio) takes two days and nights (see Fn.\ \ref{foo:rasat}); from this distance of 313\,km a speed of 156.5\,km$/$d follows.
Second, Eratosthenes probably assessed 1000\,st for a day-and-night-seafaring (see Fn.\ \ref{foo:rasat}); by means of the Egyptian stadium (cf.\ Section \ref{sec:searoutes}) a speed of 157.5\,km$/$d results.
Both results are similar and yield a length of about 940\,km for the sea route of Pytheas' voyage, which is used here.
According to Pliny (NH~IV.30) one traveled from the island Berrice (also named Nerigos in the manuscripts) to Thule.
Berrice is possibly the island Mainland of Shetland, cf.\ \cite[p.\ 61]{nan11} and \cite[p.\ 156]{hen44}.
The starting point of the six-day journey to Thule, however, was rather located at Great Britain, since Thule ``\ldots\ is six days' sail from the north of Britain \ldots'' (\cite{bos55}) according to NH~II.77 (\cite[p.\ 136]{dil85}, for example, chooses Cape Wrath for the starting point).
NH~IV.30 suggests that Berrice/Mainland of Shetland was on Pytheas' way to Thule, which is taken into account here.
For the starting point of the time measurement of the six-days journey Duncansby Head is assumed here.
Figure \ref{fig:reise} shows two possible sea routes with a length of 940\,km from Great Britain to Thule.
Both routes bypass Orkney and Mainland.
From there, route A takes course directly to the West Cape of Norway and continues alongside the Norwegian coast up into the Trondheimsfjord.
Route B takes course eastwards along a constant latitude to the Norwegian coast at Bergen and continues alongside the coast up to the island Sm{\o}la.
Possibly, such a latitude sailing was used, which was an easy and common method for navigation (it was used, e.g., by the Vikings later on, cf. \cite{joh94}).
\cite[p.\ 167]{hen44} rejects the similar route Orkney -- Bergen because in his opinion the eastern course contradicts the position of Thule north of Britain.
If, however, Pytheas visited a northern region of Norway and only referred the name Thule to this region, then there is no contradiction.
Owing to the uncertainty of the assumed speed and way, Pytheas' landing point can not be given precisely.
The Trondheimsfjord and the coastal region of the same latitude come into consideration.

\begin{figure}[p]
\centering
\includegraphics[width=8cm]{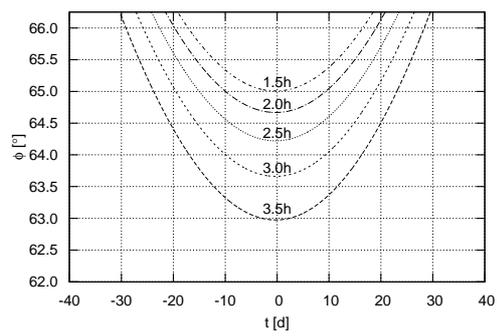}
\caption{Isolines of the length $n$ of the night subject to the latitude $\phi$ and time $t$ expressed by the number of days since the summer solstice in 330 BC}
\label{fig:nacht}
\end{figure}

Before Geminus quotes Pytheas in E~VI.9 he discusses $M$-data of different latitudes.
Thus, \cite[p.\ 57]{nan11} assumes that Pytheas' information on the length of the nights refers to the shortest nights of the year.
This, however, does not result directly from the text.
For Pytheas' time the lengths of the nights were determined; the year 330 BC was used, other supposable times do not yield significant differences.
The shortest nights (at summer solstice) with lengths $N=2\,\text{h},3\,\text{h}$ occurred at $\phi=64\degree40'$ and $\phi=63\degree40'$, respectively.\footnote{For a location of latitude $\phi$ and for a given time $t$, the length $n$ of the night can be determined by the following calculation steps (according to \cite{str12}; applied formulas see \cite[pp.\ 98, 135, 151--153]{mee91}; cf.\ also Fn.\ \ref{foo:altitude}):
$\varepsilon=\varepsilon(t)$;
$M(t)$;
$L_0=L_0(t)$;
$C=C(t,M)$;
$\Theta=\Theta(L_0, C)$;
$\delta=\delta(\Theta, \varepsilon)$;
hour angle at sunset: $H=H(\phi, \delta, a_0)$;
$n=24\,\text{h}-2H$.
By means of the altitude $a_0=-50'$ the atmospheric refraction and the size of the sun disc are taken into account.
For the determination of $\phi$, it was varied till $n$ equaled the given value.}
These latitudes are significantly less than $\phi=66\degree16'$ of the polar circle and thus inconsistent with the information on the arctic circle (see Section \ref{sec:obliquity}), see also Figure \ref{fig:reise}.
In Geminus' quotation it is only said that the night was very short so that the given lengths of nights may refer to a date near the summer solstice.
In order to locate the associated region, the length $n$ of the night was determined for different latitudes $\phi$ and times $t$.
Figure \ref{fig:nacht} shows the result in form of isolines of $n$.
At $\phi=63\degree20'$ $n$ was $3.3\,\text{h}\approx3\,\text{h}$ five days before/after the summer solstice.
Thus, the southern limit for Pytheas' Thule can be located at this latitude, which corresponds to the latitude of the southern end of the Trondheimsfjord.
At the polar circle, at $\phi=66\degree16'$, $n$ was $\approx2\,\text{h}$ 22 days before/after the summer solstice.
Possibly, Pytheas traveled to this region at that time, where he heard about the midnight sun.

Pytheas' information on the arctic circle and Eratosthenes latitudinal data lead to the northern polar circle at $\phi=66\degree16'$ at Pytheas' time.
Pytheas' information on the journey length suggests the region at the latitude of the southern end of the Trondheimsfjord.
His information on the length of the nights leads to both of these regions.
This is not contradictory because the name Thule may refer to a region of larger extent.
Hence, Pytheas' Thule can be equated with the region of Norway west of the Scandinavian Mountains between about 63\degree20$'$ and 66\degree16$'$ latitude.
This result is in accordance with Pytheas' contact with inhabitants and his report on the cultivation of grain in Thule (G~IV.5.5).
In the said region there are spacious low-lying areas and a warm and humid climate influenced by the North Atlantic Current.
Apart from the southern regions at the Skagerrak, in Norway there are only low-lying areas with fertile clayey soils at the Trondheimsfjord (see \cite[p.\ 26, fig.\ 6]{spo08}).
According to \cite[pp.\ 7--8]{hel08} there were stable settlements and farming in Norway as far north as Tr{\o}ndelag at the beginning of the Iron Age (500 BC -- 800 AD).
Furthermore, in regions at the polar circle agriculture was introduced in the 7th and 4th century BC as revealed by radiocarbon dating based on pollen (see \cite{joh86}).

%%%%%%%%%%%%%%%%%%%%%%%%%%%%%%%%%%%%%%%%%%%%%%%%%%%%%%%%%%%%%%%%%%%%%%%%%%%%%%%
%                               Bibliography
%%%%%%%%%%%%%%%%%%%%%%%%%%%%%%%%%%%%%%%%%%%%%%%%%%%%%%%%%%%%%%%%%%%%%%%%%%%%%%%

\bibliography{biblio}

%%%%%%%%%%%%%%%%%%%%%%%%%%%%%%%%%%%%%%%%%%%%%%%%%%%%%%%%%%%%%%%%%%%%%%%%%%%%%%%
%                                 Tables
%%%%%%%%%%%%%%%%%%%%%%%%%%%%%%%%%%%%%%%%%%%%%%%%%%%%%%%%%%%%%%%%%%%%%%%%%%%%%%%

\begin{table}[hl]
\caption{Latitudinal data of the western \textit{Oikoumene} derived from the fragments (F; with the exception of F34.13 from \cite{rol10}) ascribed to Eratosthenes; $^*$ = see note}
\label{tab:eratdata}
\begin{tabular} {p{1.7cm} p{1.4cm} p{3.2cm} p{3.2cm} r}
\hline
F & Source & From & To & $b$ [st] \\
\hline
30.1, 34.12 & G\,II.5.6, II.5.9 & Cinnamon country & northern regions & $<$30,000 \\
30.2, 34.6 & G\,II.5.6, II.5.7 & equator & Cinnamon country & 8,800 \\
34.1, 35.1 & G\,II.5.7, I.4.2 & Mero\"{e} & Alexandria & ${\approx}\atop{=}$10,000 \\
34.2 & G\,II.5.7 & Mero\"{e} & Syene & 5,000 \\
34.3 & G\,II.5.7 & Cinnamon country & Mero\"{e} & $\approx$3,000 \\
34.4 & G\,II.5.7 & Cinnamon country & Syene & 8,000 \\
34.5$^*$, M6.1$^*$ & G\,II.5.7, C\,I.7 & equator & Syene & 16,800 \\
34.7 & G\,II.5.7 & equator & Alexandria & 21,800 \\
34.8 & G\,II.5.8 & Byzantium & Borysthenes & $\approx$3,800 \\
34.9 & G\,II.5.9 & Borysthenes & northern regions & 4,000 \\
34.10 & G\,II.5.9 & Rhodes & northern regions & 12,700 \\
34.11 & G\,II.5.9 & Cinnamon country & Rhodes & 16,600 \\
34.13 & G\,II.5.8 & Borysthenes & northern regions & 3,000 \\
35.2 & G\,I.4.2 & Alexandria & Hellespont & $\approx$8,100 \\
35.3, 36.3 & G\,I.4.2, II.5.42 & Hellespont & Borysthenes & 5,000 \\
35.4 & G\,I.4.2 & Borysthenes & Thule & $\approx$11,500 \\
35.5 & G\,I.4.2 & Cinnamon country (, Egyptian island, Taprobane) & Mero\"{e} & 3,400 \\
35.6 & G\,I.4.2 & Cinnamon country & Thule & 38,000 \\
36.1 & G\,II.5.42 & Mero\"{e} & Borysthenes & $\gtrsim$23,000 \\
36.2, 47.1 & G\,II.5.42, G\,II.1.3 & Mero\"{e} & Hellespont & 18,000 \\
40+41$^*$ & G\,II.1.20, NH\,II.185 & Mero\"{e} & country of the Trogodytes & 0\\
56.1 & G\,II.1.33 & Alexandria & Rhodes & $\lesssim$4,000 \\
128.1 & G\,II.5.24 & Alexandria & Rhodes & 3,750 \\
\hline
\end{tabular}
\end{table}

\clearpage

\begin{longtable}[l]
{p{1.7cm} p{1.4cm} p{2.7cm} p{2.7cm} p{1.2cm} l}
\caption{Latitudinal data derived from the fragments (F; with the exception of F62.1--71.1 from \cite{dic60}) ascribed to Hipparchus; n. = north of, s. = south of}
\label{tab:hippdata}\\
\hline
F & Source & From & To & $b$ [st] & Original  \\
\hline
\endhead
15.1	&	G\,II.1.12	&	Meroe	&	Byzantium	&	$\approx$18,000	&		 \\
43.1	&	G\,II.5.35	&	Cinnamon country	&	Mero\"{e}	&	3,000	 &		 \\
43.2, 44.1	&	G\,II.5.35, II.1.13	&	equator	&	Cinnamon country	&	 ${=}\atop{\approx}$8,800	 &		 \\
43.3	&	G\,II.5.35	&	Mero\"{e}	&	Syene	&	5,000	&		\\
46.1	&	G\,II.5.36	&	Mero\"{e}-\textit{klima}	&	Ptolemais	&	 0	&		 \\
46.2	&	G\,II.5.36	&	equator	&	Mero\"{e}-\textit{klima}	&	 11,600	&	 $M=13$\,h	 \\
46.3	&	G\,II.5.36	&	equator	&	Mero\"{e}-\textit{klima}	&	 11,600	&	 see note	 \\
47.1	&	G\,II.5.36	&	Syene	&	Berenice	&	0	&		\\
47.2	&	G\,II.5.36	&	Syene	&	country of the Trogodytes	&	0	 &		 \\
47.3	&	G\,II.5.36	&	equator	&	Syene	&	16,602	&	see note	 \\
47.4	&	G\,II.5.36	&	equator	&	Syene	&	16,800	&	 $M=13\frac{1}{2}$\,h	 \\
48.1	&	G\,II.5.38	&	Alexandria-\textit{klima}	&	Alexandria	&	 $\approx$400	 &	 \\
48.2	&	G\,II.5.38	&	Alexandria	&	Cyrene	&	0	&		\\
48.3	&	G\,II.5.38	&	equator	&	Alexandria-\textit{klima}	&	 21,400	&	 $M=14$\,h	 \\
48.4	&	G\,II.5.38	&	equator	&	Alexandria-\textit{klima}	&	 21,400	&	 see note	 \\
48.5	&	G\,II.5.38	&	Alexandria-\textit{klima}	&	Carthage	&	 1,300	&		 \\
48.6	&	G\,II.5.38	&	equator	&	Carthage	&	22,730	&	 $r_\text{e}=\frac{11}{7}$	 \\
49.1	&	G\,II.5.35	&	Ptolemais (Phoenicia)	&	Sidon/Tyre	&	0	 &		 \\
49.2	&	G\,II.5.35	&	equator	&	Ptolemais (Phoenicia)	&	23,400	 &	 $M=14\frac{1}{4}$\,h	 \\
49.3	&	G\,II.5.35	&	Alexandria	&	Ptolemais (Phoenicia)	&	 $\approx$1,600	 &		 \\
49.4	&	G\,II.5.35	&	Carthage	&	Ptolemais (Phoenicia)	&	 $\approx$700	 &	  \\
50.1	&	G\,II.5.39	&	Rhodes	&	Peloponnese	&	0	&		\\
50.2	&	G\,II.5.39	&	Rhodes	&	Xanthus	&	0	&		\\
50.3	&	G\,II.5.39	&	Rhodes	&	Syracuse	&	400	&		\\
50.4	&	G\,II.5.39	&	equator	&	Rhodes	&	25,400	&	 $M=14\frac{1}{2}$\,h	 \\
50.5	&	G\,II.5.39	&	Alexandria	&	Rhodes	&	3,640	&		\\
51.1	&	G\,II.5.40	&	Alexandria in the Troad	&	Amphipolis	&	0	 &		 \\
51.2	&	G\,II.5.40	&	Alexandria in the Troad	&	Apollonia	&	0	 &		 \\
51.3	&	G\,II.5.40	&	Alexandria in the Troad	&	s. Rome \& n. Naples	 &	0	 &		 \\
51.4	&	G\,II.5.40	&	equator	&	Alexandria in the Troad	&	28,800	 &	 $M=15$\,h	 \\
51.5	&	G\,II.5.40	&	Alexandria	&	Alexandria in the Troad	&	 7,000	&		 \\
51.6	&	G\,II.5.40	&	equator	&	Alexandria in the Troad	&	 $>$28,800	&		 \\
51.7	&	G\,II.5.40	&	Rhodes	&	Alexandria in the Troad	&	3,400	 &		 \\
51.8	&	G\,II.5.40	&	Alexandria in the Troad	&	Byzantium	&	 1,500	&		 \\
51.9	&	G\,II.5.40	&	Byzantium	&	Nicaea	&	0	&		\\
51.10, 53.2, 54.1, 55.1	&	G\,II.5.40, I.4.4, II.5.8, II.1.12	&	 Byzantium	&	 Massalia	&	 0	 &		 \\
52.1	&	G\,II.5.41	&	equator	&	Byzantium	&	30,300	&	 $M=15\frac{1}{4}$\,h	 \\
52.2	&	G\,II.5.41	&	equator	&	Byzantium	&	30,243	&	 $r_\text{s}=120/41\frac{4}{5}$	 \\
52.3	&	G\,II.5.41	&	Rhodes	&	Byzantium	&	$\approx$4,900	&		 \\
52.4	&	G\,II.5.41	&	equator	&	Byzantium	&	$\approx$30,300	&		 \\
53.1	&	G\,I.4.4	&	Borysthenes	&	Britain(?)	&	0	&		\\
56.1	&	G\,II.5.41	&	Byzantium	&	Mid-Pontus	&	$\approx$1,400	 &		 \\
56.2	&	G\,II.5.41	&	equator	&	Mid-Pontus	&	31,700	&	 $M=15\frac{1}{2}$\,h	 \\
56.3	&	G\,II.5.41	&	equator	&	Mid-Pontus	&	31,500	&	see note	 \\
57.1	&	G\,II.5.42	&	Byzantium	&	Lk. Maeotis	&	$\approx$3,800	 &		 \\
57.2	&	G\,II.5.42	&	equator	&	Lk. Maeotis	&	34,100	&	 $M=16$\,h	\\
57.3	&	G\,II.5.42	&	equator	&	Lk. Maeotis	&	34,100	&		\\
57.4	&	G\,II.5.42	&	equator	&	Lk. Maeotis	&	33,833	&	 $a_\text{w}=9$\,c	 \\
58.1, 59.3	&	G\,II.1.18, II.1.12	&	Borysthenes	&	Celtica	&	0	&		 \\
58.2	&	G\,II.1.18	&	equator	&	Borysthenes	&	33,833	&	 $a_\text{w}=9$\,c	 \\
59.1	&	G\,II.1.12	&	Byzantium	&	Borysthenes	&	3,700	&		 \\
59.2	&	G\,II.1.12	&	Massalia	&	Borysthenes	&	3,700	&		 \\
59.4	&	G\,II.1.13	&	equator	&	Borysthenes	&	34,000	&		\\
60.1	&	G\,II.5.42	&	Byzantium	&	n. Lk. Maeotis	&	7,700	& 6,300\,st 		 \\
60.2	&	G\,II.5.42	&	equator	&	n. Lk. Maeotis	&	38,033	&	 $a_\text{w}=6$\,c	\\
60.3	&	G\,II.5.42	&	equator	&	n. Lk. Maeotis	&	38,000	&	 $M=17$\,h	 \\
61.1	&	G\,II.1.18	&	Massalia	&	Celtica	&	7,700	&	 6,300\,st	\\
61.2	&	G\,II.1.18	&	equator	&	Celtica	&	38,033	&	 $a_\text{w}=6$\,c	 \\
61.3	&	G\,II.1.18	&	Massalia	&	18\,h-region	&	10,500	& 9,100\,st 		 \\
61.4	&	G\,II.1.18	&	equator	&	18\,h-region	&	40,833	&	 $a_\text{w}=4$\,c	\\
61.5	&	G\,II.1.18	&	equator	&	inhabited region	&	$>$42,233	 &	 $a_\text{w}<3$\,c	 \\
61.6	&	G\,II.1.18	&	equator	&	inhabited region	&	42,800	&	 $M=19$\,h	 \\
61.7	&	G\,II.1.18	&	equator	&	18\,h-region	&	40,800	&	 $M=18$\,h	 \\
62.1	&	CP\,1.3.6	&	equator	&	Greece	&	25,809	&	 $r_\text{e}=\frac{4}{3}$	 \\
62.2	&	CP\,1.3.6	&	equator	&	Greece	&	26,024	&	 $M=14\frac{3}{5}$\,h	 \\
62.3	&	CP\,1.3.6	&	equator	&	Greece	&	$\approx$25,900	&	 $a_\text{p}\approx37\degree$	 \\
63.1	&	CP\,1.3.7	&	equator	&	Alexandria in the Troad	&	28,753	 &	 $M=15$\,h	\\
63.2	&	CP\,1.3.7	&	equator	&	Alexandria in the Troad	&	 $\approx$28,700	&	 $a_\text{p}\approx41\degree$	\\
64.1	&	CP\,I.3.12	&	equator	&	Athens	&	26,024	&	 $M=14\frac{3}{5}$\,h	 \\
64.2	&	CP\,I.3.12	&	equator	&	Athens	&	$\approx$25,900	&	 $a_\text{p}\approx37\degree$	 \\
65.1	&	CP\,I.4.8	&	equator	&	Athens	&	25,809	&	 $r_\text{e}=\frac{4}{3}$	 \\
65.2, 70.1	&	CP\,I.4.8, I.11.8	&	equator	&	Athens	&	 $\approx$25,900	 &	 $\zeta_\text{a} \approx 37\degree$	 \\
66.1, 67.1, 71.1	&	CP\,I.7.11, I.7.14, II.4.2	&	equator	&	Greece	 &	 25,308	 &	 $M=14\frac{1}{2}$\,h	 \\
68.1	&	CP\,I.7.21	&	equator	&	Athens	&	25,900	&	 $\zeta_\text{a} = 37\degree$	 \\
69.1	&	CP\,I.11.2	&	equator	&	Greece	&	25,900	&	 $\zeta_\text{a} = 37\degree$	 \\
70.2	&	CP\,I.11.8	&	equator	&	Rhodes	&	25,200	&	 $\zeta_\text{a} = 36\degree$	 \\
\hline
\end{longtable}

%%%%%%%%%%%%%%%%%%%%%%%%%%%%%%%%%%%%%%%%%%%%%%%%%%%%%%%%%%%%%%%%%%%%%%%%%%%%%%%
%                                 Figures
%%%%%%%%%%%%%%%%%%%%%%%%%%%%%%%%%%%%%%%%%%%%%%%%%%%%%%%%%%%%%%%%%%%%%%%%%%%%%%%

\clearpage

\end{document}